\documentclass[prx,reprint,twocolumn,showpacs,superscriptaddress]{revtex4-2}
\usepackage[export]{adjustbox}
\usepackage{amsmath,amssymb,bm,mathrsfs,graphicx, braket, times, mathtools}
\usepackage[colorlinks=true,citecolor=blue,linkcolor=blue]{hyperref}
\usepackage{multirow}
\usepackage[all,cmtip]{xy}
\usepackage[normalem]{ulem}
\usepackage[usenames,dvipsnames]{color}
\usepackage{tikz-cd}
\usepackage[compat=1.0.0]{tikz-feynman}
\usepackage{bbding}
\usepackage{graphicx}
\usepackage{subfigure}

\usepackage[utf8]{inputenc}
\usepackage[english]{babel}
\usepackage{amsthm}
\usepackage{wasysym}

\renewcommand{\vec}[1]{\mathbf{#1}}

\begin{document}

\title{Stabilizer Perturbation Theory: A Systematic Construction via Schrieffer-Wolff Transformation}

\author{Xuzhe Ying} \email{yingxz@ust.hk}\thanks{These authors contributed equally.}
\affiliation{Department of Physics, Hong Kong University of Science and Technology, Clear Water Bay, Hong Kong, China}
\author{Kangle Li}\thanks{These authors contributed equally.}
\affiliation{Department of Physics, Hong Kong University of Science and Technology, Clear Water Bay, Hong Kong, China}
\affiliation{Department of Physics, National University of Singapore, Singapore 117542}
\author{Hoi Chun Po} \email{hcpo@ust.hk}
\affiliation{Department of Physics, Hong Kong University of Science and Technology, Clear Water Bay, Hong Kong, China}

\begin{abstract}
Perturbation theories provide valuable insights on quantum many-body systems. Systems of interacting particles, like electrons, are often treated perturbatively around exactly solvable Gaussian points. 
Systems of interacting qubits have gained increasing prominence as another class of models for quantum systems thanks to the recent advances in experimentally realizing mesoscopic quantum devices. 
Stabilizer states, innately defined on systems of qudits, have correspondingly emerged as another class of classically simulatable starting point for the study of quantum error-correcting codes and topological phases of matter in such devices. As a step towards analyzing more general quantum many-body problems on these platforms, we develop a systematic stabilizer perturbation theory in qubit systems. Our approach relies on the local Schrieffer-Wolff transformation, which we show can be efficiently performed through the binary encoding the Pauli algebra. As demonstrations, we first benchmark the stabilizer perturbation theory on the transverse field Ising chain in one dimension. The method is then further applied to $\mathbb{Z}_2$ toric code on square lattice and kagome lattice to probe the tendency toward confinement for anyonic excitations. 
\end{abstract}

\maketitle
\tableofcontents

\section{Introduction}

Stabilizer states is a particular set of quantum states, which are extremely useful for quantum error correction \cite{PhysRevA.52.R2493,PhysRevA.57.127,nielsen2010quantum}. A wide variety of fault-tolerant codes falls into the category of stabilizer code. Examples include Shor code \cite{PhysRevA.52.R2493}, Calderbank-Shor-Steane code \cite{PhysRevA.54.1098,Steane1996QEC}, toric code \cite{Kitaev1997QECwImperfectGates,kitaev1997quantum,KITAEV20032}, and surface codes \cite{bravyi1998quantum,freedman2001projective,dennis2022TopoQuantumMemory,PhysRevA.86.032324}. Some of the stabilizer states exhibit topological order and are robust against local perturbations \cite{bravyi2010TopoQuantumOrder,bravyi2011shortproof,michalakis2013StabilityFFHamiltonian}.

The well-known Gottesman-Knill theorem states that stabilizer states can be efficiently simulated classically \cite{gottesman1998heisenberg,PhysRevA.70.052328}. In other words, the set of stabilizer states lacks the necessary quantum resource for a universal quantum computing, namely the ``magic'' \cite{veitch2014resourceStab} - either a non-Clifford gate (such as $\pi/8$-gate) \cite{gottesman1998heisenberg,nielsen2010quantum} or some perturbations from stabilizer state (``magic state'') \cite{veitch2014resourceStab,doi:10.1142/S0217979213450197} is required for universal quantum computing. 
In addition, there is always perturbation of some sort in reality. Hence, simulating a general state with ``magic'' is an important task. In this work, we aim at a systematic perturbative description of perturbed stabilizer states, where dealing with non-Clifford unitary is inevitable.

A closely related situation is the (fermionic) gaussian states and matchgates \cite{10.1145/380752.380785,PhysRevA.65.032325,jozsa2008matchgates}. The gaussian states correspond to free theories. Adding interactions (e.g., quartic terms in fermionic operators) to the free Hamiltonian greatly enhances the difficulties in simulations. An error-free simulation requires exponential classical computational resources. Nevertheless, many important features of interacting systems can be deduced from a perturbative analysis. The most famous example is the Landau Fermi liquid theory \cite{landau1957theory,landau1959theory}. Indeed, the Wick's theorem of a gaussian state warrants a systematic perturbative analysis based on the diagrammatic technique \cite{abrikosov2012methods,mahan2013many}. In many cases, the diagrammatic perturbative technique can even be applied to non-equilibrium systems \cite{stefanucci2013nonequilibrium,kamenev2023field}.

For a generic many-body Hamiltonian, perturbations can be done with a technique called Schrieffer-Wolff (SW) transformation \cite{PhysRev.149.491}. One of the most well-known success of SW transformation lies in the application to the Kondo effect \cite{PhysRev.149.491}, where J. R. Schrieffer and P. A. Wolff applied a canonical transformation, now referred to as Schrieffer-Wolff transformation, to relate the Anderson impurity model \cite{PhysRev.124.41} to the s-d exchange model of Kondo problem \cite{Kondo1964ResisMinimum}. An infinitesimal version of SW transformation, known as flow equations, was developed in Refs.~\cite{PhysRevD.48.5863,https://doi.org/10.1002/andp.19945060203,WEGNER200177} and then evolve into a systematic calculation scheme, termed perturbative continuous unitary transformation \cite{stein1997FlowEq,knetter2000PertTheory,knetter2003StructureOp}. The idea of flow equation was later employed to prove the stability of quantum topological order \cite{bravyi2010TopoQuantumOrder,bravyi2011shortproof,michalakis2013StabilityFFHamiltonian}. Recently, the SW transformation was applied to study the properties of interacting narrow bands as is the case in twisted bilayer graphene \cite{doi:10.1073/pnas.2217816120,PhysRevB.109.024507}. A recent technical review on the Schrieffer-Wolff transformation can be found in Ref.~\cite{BRAVYI20112793}.

In this work, we develop a systematic SW transformation to describe perturbed stabilizer states, termed as the stabilizer perturbation theory. We explicitly construct the SW transformation in a perturbative way. The construction relies on the stabilizer nature of the unperturbed state and Pauli algebra, which allows straightforward numerical implementation. The stabilizer perturbation theory we develop in this paper has additional advantages, including locality preserving and symmetry preserving. We should mention that throughout the paper, we \emph{assume} the unperturbed stabilizer state is stable under perturbations to be introduced. Namely, the energy gap between the ground states and the excited states is assumed to be finite under weak perturbation.

We apply the stabilizer perturbation to a few models, including transverse Ising chain in one dimension and toric code models on square lattice and kagome lattice in two dimensions. For transverse field Ising chain in one dimension, physical quantities, such as magnetization and static correlation functions, are calculated. The perturbative results exhibit good match with density matrix renormalization group (DMRG). Moreover, the correlation length estimated from perturbatively computed correlation function exhibits a diverging tendency towards the critical point. Hence, the perturbative calculation is sensitive to the presence of a phase transition point. For toric code models, we calculate the expectation value of loop operators to probe the tendency toward confinement for some anyons. We study various types of perturbations, such as transverse field, interlayer Ising coupling for a bilayer system as well as the nearest-neighbor Ising and Heisenberg coupling for toric code model on Kagome lattice. In the perturbative calculation, the deviation from perimeter law provides good estimation of the critical point out of the toric code state.  

The rest of the article is organized as follows:
\begin{itemize}
    \item Section~\ref{Sec:Preliminary} reviews the basics of Pauli group and the stabilizer states. In this section, we introduce notations and conventions for later discussions. 
    \item Section~\ref{Sec:StabHam_HilbertSpace} introduces the context, where the stabilizer state is considered as the ground state of certain (unperturbed) stabilizer Hamiltonian. 
    \item Sections~\ref{Sec:PertStabHam_LocalSWTransformation}-\ref{Sec:NumericalImplementation} present the stabilizer perturbation theory on the analytical construction and numerical implementation separately. 
    \item Sections~\ref{Sec:Application:TFIC}-\ref{Sec:Application_TC_kagome} apply the stabilizer perturbation method to a few examples: (1) Transverse field Ising chain in 1D in Section~\ref{Sec:Application:TFIC}; (2) Toric code on square lattice in Section~\ref{Sec:Application_TC_squarelattice}; (3) Toric code on kagome lattice in Section~\ref{Sec:Application_TC_kagome}.
    \item Section~\ref{Sec:Conclusion_discussion} concludes this article with further discussions. 
\end{itemize}    

This article is appended with an Appendix for certain technical details. Appendix~\ref{Sec:PauliAlgebra_Implementation} outlines the numerical implementation of Pauli algebra arithmetic. Appendix~\ref{Sec:Verify_DeStab_CM} verifies the construction of de-stabilizer operators. 

\section{Preliminary: Pauli Group and Stabilizer States}
\label{Sec:Preliminary}

In this paper, we consider qubit systems. To begin with, we review basics on Pauli group and stabilizer states. This will fix the nomenclatures, notations, and conventions for later discussions on stabilizer perturbation theory.

\subsection{Pauli Group}

Throughout the paper, we fix the Pauli matrices to take the following form:
\begin{equation}
    \begin{split}
        &I=\begin{bmatrix}
            1 & 0\\
            0 & 1
        \end{bmatrix},\ X=\begin{bmatrix}
            0 & 1\\
            1 & 0
        \end{bmatrix},\ 
        Y=\begin{bmatrix}
            0 & -i\\
            i & 0
        \end{bmatrix},\ Z=\begin{bmatrix}
            1 & 0\\
            0 & -1
        \end{bmatrix}.
    \end{split}
\end{equation}

Consider a system of $N$ qubits. The $N$-qubit Pauli group $P_N$ is generated by $\langle i,X,Z\rangle^{\otimes N}$. Namely, the elements of the Pauli group is a tensor products of $\{I,X,Y,Z\}$ together with an overall factor of $\{\pm1, \pm i\}$. We call a generic element in $P_N$ a Pauli operator. We call the element $P_s\in P_N$ a Pauli string, if its prefactor is $+1$, i.e., $P_s$ is a direct tensor product of Pauli matrices. A generic $N$-qubit operator can be expressed as a superposition of Pauli strings: $\hat{O}=\sum c_{P_s}P_s$. We allow the superposition coefficients $c_{P_s}$ to be complex, such that both Hermitian operators and anti-Hermitian operators are included in the description. The generic $N$-qubit operators together with operator arithmetic is called Pauli algebra. 

The Pauli operators (as well as a general operator $\hat{O}=\sum c_{P_s}P_s$) can be encoded in binary arrays \cite{PhysRevA.70.052328,nielsen2010quantum,PhysRevA.68.042318,10.21468/SciPostPhysCodeb.54,10.21468/SciPostPhysCodeb.54-r1.5}. The operator arithmetic can be carried out efficiently with the binary encoding in numerics. 

To see how the encoding works, let us begin with single qubit case. The single qubit Pauli matrices can be concisely expressed as Heisenberg-Weyl operators, characterized by two $\mathbb{Z}_2$-valued variables $(a,b)\in \mathbb{Z}_2\otimes\mathbb{Z}_2$:
\begin{equation}
    T^{(a,b)}=e^{i\frac{\pi}{2}ab}X^aZ^b
\end{equation}
It is straightforward to verify that $I=T^{(0,0)}$, $X=T^{(1,0)}$, $Y=T^{(1,1)}$, and $Z=T^{(0,1)}$. Hence, each Pauli matrix can be represented by two $\mathbb{Z}_2$ variables. 

A generic single qubit operator can be parameterized as:
\begin{equation}
    \hat{O}=\sum_{i=1}^{N_{\text{op}}}c_iT^{(a_i,b_i)}
\end{equation}
where $N_{\text{op}}$ is the total number of Pauli matrices to be summed over. Numerically, this operator $\hat{O}$ is represented by two matrices/arrays: the check matrix and the coefficient vector: 
\begin{equation}
    \mathsf{CM}_{\hat{O}}:=\begin{bmatrix}
        a_1 & b_1\\
        a_2 & b_2\\
        \vdots & \vdots\\
        a_{N_{\text{op}}} & b_{N_{\text{op}}}
    \end{bmatrix}, \ \ \ \ \mathsf{Coeff}_{\hat{O}}=\begin{bmatrix}
        c_1 \\ c_2 \\ \vdots \\ c_{N_{\text{op}}}
    \end{bmatrix}.
\end{equation}
The first matrix $\mathsf{CM}_{\hat{O}}$ is called the check matrix, storing the information of Pauli matrices. Namely, the $i$-th row consists of the $\mathbb{Z}_2$ variables $(a_i,b_i)$ for the Pauli matrix $T^{(a_i,b_i)}$. The second matrix $\mathsf{Coeff}_{\hat{O}}$ stores the additive coefficients $c_i$, with caution that the order has to be consistent with the order of the rows in the check matrix.

Similarly, for an $N$-qubit system, a Pauli string can be represented by two vectors $(\vec{a},\vec{b})\in \mathbb{Z}_2^{\otimes N}\otimes\mathbb{Z}_2^{\otimes N}$:
\begin{equation}
    T^{(\vec{a},\vec{b})}=\otimes_{j=1}^{N}T^{(a_j,b_j)} 
\end{equation}
where $\vec{a}:=\left[a_1,a_2,\cdots,a_N\right]$ and $\vec{b}:=\left[b_1,b_2,\cdots,b_N\right]$; $T^{(a_j,b_j)}$ is the Pauli matrix acting on the $j$-th qubit. A generic $N$-qubit operator
\begin{equation}
    \hat{O}:=\sum_{i=1}^{N_{\text{op}}}c_iT^{(\vec{a}_i,\vec{b}_i)}
\end{equation}
can be represented by the check matrix and coefficient vector as well: 
\begin{equation}
    \mathsf{CM}_{\hat{O}}=\begin{bmatrix}
        \vec{a}_1 & \vec{b}_1\\
        \vec{a}_2 & \vec{b}_2\\
        \vdots & \vdots\\
        \vec{a}_{N_{\text{op}}} & \vec{b}_{N_{\text{op}}}
    \end{bmatrix},\ \ \ \ \mathsf{Coeff}_{\hat{O}}=\begin{bmatrix}
        c_1 \\ c_2 \\ \vdots \\ c_{N_{\text{op}}}
    \end{bmatrix}.
\end{equation}
The check matrix $\mathsf{CM}_{\hat{O}}$ is an $N_{\text{op}}\times 2N$ $\mathbb{Z}_2$ valued matrix, encoding the information of Pauli strings in $\hat{O}$. Accordingly, the coefficients are stored in $\mathsf{Coeff}_{\hat{O}}$, with the order consistent with the order of the rows in the check matrix.

The Pauli algebra arithmetic can be implemented based on the check matrix and coefficient vector and our implementation is outlined in Appendix~\ref{Sec:PauliAlgebra_Implementation}. Similar implementations can be found in the literature, such as Refs.~\cite{PhysRevA.68.042318,10.21468/SciPostPhysCodeb.54,10.21468/SciPostPhysCodeb.54-r1.5}.

Let us specify two operations, multiplication and commutation check. First, multiplication of two Pauli strings is given by:
\begin{equation}
    T^{(\vec{a}_1,\vec{b}_1)}T^{(\vec{a}_2,\vec{b}_2)}=F\left(\vec{a}_1,\vec{b}_1,\vec{a}_2,\vec{b}_2\right)T^{(\vec{a}_1\oplus\vec{a}_2,\vec{b}_1\oplus\vec{b}_2)}
\end{equation}
where for brevity $\vec{a}\oplus\vec{b}:=\vec{a}+\vec{b}\mod2$. The overall phase is given by:
\begin{equation}
    \begin{split}
        &F\left(\vec{a}_1,\vec{b}_1,\vec{a}_2,\vec{b}_2\right)
        :=F\left(\mathsf{CM}_1,\mathsf{CM}_2\right)\\
        =&e^{i\frac{\pi}{2}\left(\vec{a}_2\cdot\vec{b}_1-\vec{a}_1\cdot\vec{b}_2\right)}
        \times e^{i\frac{\pi}{2}\left[\left(\vec{a}_1+\vec{a}_2\right)\cdot\left(\vec{b}_1+\vec{b}_2\right)-\left(\vec{a}_1\oplus\vec{a}_2\right)\cdot\left(\vec{b}_1\oplus\vec{b}_2\right)\right]}
    \end{split}
    \label{Eq:PauliProdCoeff}
\end{equation}
where $\mathsf{CM}_i=\left[\vec{a}_i\quad\vec{b}_i\right]$; the factor of $e^{i\frac{\pi}{2}\left(\vec{a}_2\cdot\vec{b}_1-\vec{a}_1\cdot\vec{b}_2\right)}$ on the second line is sensitive to the order of operators to be multiplied. It is straightforward to show that 
\begin{equation}
    T^{(\vec{a}_1,\vec{b}_1)}T^{(\vec{a}_2,\vec{b}_2)}=e^{i\pi\left(\vec{a}_2\cdot\vec{b}_1-\vec{a}_1\cdot\vec{b}_2\right)}T^{(\vec{a}_2,\vec{b}_2)}T^{(\vec{a}_1,\vec{b}_1)}.
\end{equation}
Then, one can deduce the operation of checking commutation relation. One immediate implication is that any two Pauli strings either commute or anti-commute. This property simplifies the numerical implementations significantly.

Second, checking if two Pauli strings commute can be achieved by computing:
\begin{equation}
\begin{split}
    &\vec{a}_2\cdot\vec{b}_1\oplus\vec{a}_1\cdot\vec{b}_2
    =\left\{\begin{split}
        & 0\ \ \ \text{for}\ \left[T^{(\vec{a}_1,\vec{b}_1)},T^{(\vec{a}_2,\vec{b}_2)}\right]=0\\
        & 1\ \ \ \text{for}\ \left[T^{(\vec{a}_1,\vec{b}_1)},T^{(\vec{a}_2,\vec{b}_2)}\right]\neq0
    \end{split}\right.
\end{split}
\end{equation}
where we used the fact that $\left(\vec{a}_2\cdot\vec{b}_1-\vec{a}_1\cdot\vec{b}_2\right)\mod2=\vec{a}_2\cdot\vec{b}_1\oplus\vec{a}_1\cdot\vec{b}_2$. 

More generally, one can employ the check matrix to check the commutation relation between two operators $\hat{O}_{1,2}$ with the check matrices $\mathsf{CM}_{\hat{O}_{1,2}}$:
\begin{equation}
    M_{[\hat{O}_1,\hat{O}_2]}:=\mathsf{CM}_{\hat{O}_1}\ \hat{L}\ \mathsf{CM}_{\hat{O}_2}^{\text{T}}\mod2
\end{equation}
where $\hat{L}$ is a $2N\times 2N$ matrix:
\begin{equation}
    \hat{L}=\begin{bmatrix}
    0_{N\times N} & \mathbb{I}_{N\times N}\\
    \mathbb{I}_{N\times N} & 0_{N\times N}
\end{bmatrix}
\end{equation}
The matrix element $M_{[\hat{O}_1,\hat{O}_2]}^{ij}=0$ (or $1$) dictates that the $i$-th Pauli string in $\hat{O}_1$ and the $j$-th Pauli string in $\hat{O}_2$ commute (or anti-commute).

\subsection{Stabilizer States}
\label{Sec:StabilizerState_Review}

Now consider a multi-qubit stabilizer state, which by definition is the common eigenstate of a set of commuting Pauli operators with \emph{all} the eigenvalues equal to $+1$. The stabilizer group is generated by a set of commuting Pauli operators 
\begin{equation}
\begin{split}
    &\mathsf{Stab}=\langle g_1,g_2,\cdots, g_l\rangle,\ \text{with}\ [g_i,g_j]=0,\ \forall\ i,j
\end{split}
\end{equation}
with $l\leq N$ and $g_i$'s are independent. The group elements ($s\in \mathsf{Stab}$) of the stabilizer group has a general structure:
\begin{equation}
    s=\prod_{i=1}^lg_i^{n_{s,i}},\ n_{s,i}=0,1
    \label{Eq:StabGroupElements}
\end{equation}
A stabilizer state $|0\rangle$ is defined as:
\begin{equation}
    g_i|0\rangle=|0\rangle,\ \forall\ i
    \label{Eq:Def_StabState}
\end{equation}
Namely, $|0\rangle$ is a common eigenstate of all the generators of the stabilizer group, with eigenvalues equal to identity. It is straightforward to see that the stabilizer state $|0\rangle$ is also an eigenstate of all group elements in $\mathsf{Stab}$, namely $s|0\rangle=|0\rangle$ for all $s\in\mathsf{Stab}$. Notice that the states $|0\rangle$ satisfying the definition above is not unique if $l<N$. Indeed, the collection of the stabilizer states, stabilized by the operators $\langle g_1,g_2,\cdots,g_l\rangle$, spans a Hilbert subspace of dimension $2^{N-l}$.

The Gottesman-Knill theorem states that stabilizer states can be efficiently simulated classically \cite{gottesman1998heisenberg,PhysRevA.70.052328}. Indeed, different stabilizer states can be transformed into each other by a Clifford circuit. The construction of Clifford circuit was shown to be in the $\oplus\mathsf{L}(\subseteq\mathsf{P})$ complexity class \cite{PhysRevA.70.052328}, where only polynomially many $\mathsf{NOT}$ and $\mathsf{CNOT}$ gates/circuits is required. It turns out that a small number of non-Clifford gates is also simulatable, with required computational resource scales exponentially with the number of non-Clifford gates \cite{PhysRevA.70.052328}.

\section{Stabilizer Hamiltonian and Hilbert Space Structure}
\label{Sec:StabHam_HilbertSpace}

In the rest of the paper, we consider the stabilizer states as the ground state(s) of a stabilizer Hamiltonian (denoted as $H_0$, namely the unperturbed Hamiltonian):
\begin{equation}
    H_0=-\sum_jh_js_j,
\end{equation}
where $s_j\in P_N$ are a set of \emph{commuting} Pauli operators ($[s_j,s_{j^{\prime}}]=0$ for all $j,j^{\prime}$) with positive additive coefficients $h_j>0$. Suppose the Pauli operators $s_j$ are properly chosen such that the Hamiltonian $H_0$ is frustration free. Then, the ground state is stabilized by all the Pauli operators $s_j$ in $H_0$, i.e., an eigenstate of all the $s_j$ with eigenvalue equal to $+1$.

Suppose $\{g_1,g_2,\cdots,g_l\}$ is the set of independent elements of $\{s_j\}$. One can define the stabilizer group as $\mathsf{Stab}=\langle g_1,g_2,\cdots,g_l\rangle$ and correspondingly $s_j\in\mathsf{Stab}$.

As mentioned briefly, the ground state subpsace is $2^{N-l}$-fold degenerate ($N$ is the number of qubits). The projector onto the ground-state subspace can be conveniently constructed by the stabilizer operators:
\begin{equation}
    \hat{P}_0=\prod_{i=1}^l\frac{1}{2}\left(1+g_i\right)=\frac{1}{2^l}\sum_{\forall s\in\mathsf{Stab}}s
    \label{Eq:GSProjector}
\end{equation}
The grounstate energy is given by $E_0=-\sum_ih_i$. 

Without ambiguity, we use $|0\rangle$ to denote any state in the ground-state subspace. In certain cases, we might want to specify a particular state in the ground-state subspace. This is done by specifying $(N-l)$ additional stabilizer operators $\{\tilde{g}_{l+1},\tilde{g}_{l+2},\cdots,\tilde{g}_N\}$ with $\tilde{g}_i=(-1)^{n_i}\tilde{Z}_i$ and $n_i=0,1$ for $i=l+1,\cdots,N$. Here, $\tilde{Z}_i$ are some Pauli strings. The choice of the signs $(-1)^{n_i}$ dictates the $2^{N-l}$-fold degeneracy of the ground-state subspace. In a more quantum computational language, the ground-state subspace defines $(N-l)$ logical qubits. $\tilde{Z}_i$ can be considered as the \emph{logical} Pauli $Z$-operators. Correspondingly, the states stabilized by $\{g_1,\cdots, g_l,\tilde{g}_{l+1},\cdots,\tilde{g}_N\}$ define the \emph{logical} computational basis \cite{nielsen2010quantum}.

The Hilbert space is divided into subspaces of dimension $2^{N-l}$. Indeed, the projector onto certain subspace $n=\{n_i,i=1,2,\cdots,l\}$ is given by:
\begin{equation}
    \hat{P}_n=\prod_{i=1}^l\frac{1}{2}\left[1+\left(-1\right)^{n_i}g_i\right]
\end{equation}
where $n_i=0,1$ for all $i=1,2,\cdots,l$. This subspace is an eigenspace of all stabilizer operators $\{g_1,g_2,\cdots,g_l\}$ with eigenvalues $\{(-1)^{n_1},(-1)^{n_2},\cdots,(-1)^{n_l}\}$. The energy of this subspace is given by:
\begin{equation}
    E_n=-\sum_j h_j\ (-1)^{\sum_{i=1}^{l}n_in_{s_j,i}}
\end{equation}
where $s_j=\prod_{i=1}^lg_i^{n_{s_j,i}}$ is used [Eq.~(\ref{Eq:StabGroupElements})]. Or equivalently:
\begin{equation}
    E_n-E_0=-\sum_j h_j\ \left[(-1)^{\sum_{i=1}^{l}n_in_{s_j,i}}-1\right]
    \label{Eq:EnergyDiff_n&0}
\end{equation}
Namely, the energy difference between the excited subspace ``$n$'' and the ground-state subspace ``$0$'' is determined by the Pauli operators $s_j$ in $H_0$, whose eigenvalue on the excited subspace is $-1$. An immediate implication is that there is a finite energy gap between ground-state subspace and all the excited energy states. 

When there is no ambiguity, we use $|n\rangle$ to denote any state in this subspace $n$. To specify a particular state in this subspace, one can use the same set of additional stabilizer operators $\{\tilde{g}_{l+1},\tilde{g}_{l+2},\cdots,\tilde{g}_N\}$ as the case of ground-state subspace.

\section{Perturbing Stabilizer Hamiltonian: Local Schrieffer-Wolff Transformation}
\label{Sec:PertStabHam_LocalSWTransformation}

In this section, we add some perturbation $H_1$ to the stabilizer Hamiltonian $H_0$ and introduce the stabilizer perturbation theory based on the local Schrieffer-Wolff transformation. Throughout the paper, we \emph{assume} the unperturbed stabilizer Hamiltonian is stable under perturbation. Namely, the energy gap between the ground states and the excited states is assumed to be finite under weak perturbation.

The perturbative calculation below employs the Pauli algebra and stabilizer nature of the unperturbed state. Hence, it is suitable for numerical implementation. In this section, we introduce the analytical construction for the stabilizer perturbation theory. In Section~\ref{Sec:NumericalImplementation}, we outline the numerical implementation.

\subsection{Adding Perturbations}
Consider a general perturbation:
\begin{equation}
    H_1=\sum_j\lambda_j\mathcal{O}_j
\end{equation}
where $\mathcal{O}_j\in P_N$. Given the stability assumption of the unperturbed stabilizer state, we ask how the ground states and their properties are modified under perturbation.

Now the full Hamiltonian is given by $H=H_0+H_1$
with the perturbed ground state denoted by $|\text{G.S.}\rangle$ (this is understood as some state in the ground-state subspace if there is degeneracy). The stability assumption implies that the perturbed ground state and the unperturbed ground state belongs to the same phase $|\text{G.S.}\rangle\sim|0\rangle$. There exists a unitary such that $|\text{G.S.}\rangle=U|0\rangle$. Generically, the unitary $U$ can be viewed as a (quasi-)adiabatic evolution from the unperturbed state $|0\rangle$ to the perturbed state $|\text{G.S.}\rangle$ \cite{PhysRevB.72.045141,PhysRevB.82.155138}.
Ref.~\cite{PhysRevB.82.155138} further suggests that the unitary $U$ can be expressed as a constant depth quantum circuit. Below, we construct the unitary $U$ in a perturbative way via the method of local Schrieffer-Wolff transformation \cite{BRAVYI20112793,Datta1996localSWTransformation}.

The method of local Schrieffer-Wolff transformation perturbatively constructs the unitary transformation $U=e^{S}$ in a way that the Hamiltonian 
\begin{equation}
    \tilde{H}=e^{-S}\left(H_0+H_1\right)e^S
    \label{Eq:Transformed_Hamiltonian}
\end{equation}
becomes approximately block-diagonal. Namely, the unperturbed ground state is an approximate eigenstate: $\tilde{H}|0\rangle\approx \tilde{E}_0|0\rangle$. The anti-Hermitian operator $S=S^{(1)}+S^{(2)}+\cdots$ is constructed perturbatively. Then, the ground state property, such as expectation value of certain operator $\hat{O}$, is evaluated by:
\begin{equation}
    \langle\text{G.S.}|\hat{O}|\text{G.S}\rangle\approx\langle 0|e^{-S}\hat{O}e^{S}|0\rangle
\end{equation}
It turns out that the local Schrieffer-Wolff transformation can be conveniently constructed based on Pauli algebra arithmetic. Thus, numerical computation becomes feasible. In addition, if one takes advantage of translation invariance, the computation can be carried out for quite large systems. More details are as follows.

\subsection{Local Schrieffer-Wolff (SW) Transformation}

As briefly mentioned above, the local SW transformation aims to construct a new Hamiltonian $\tilde{H}=e^{-S}\left(H_0+H_1\right)e^{S}$ that is approximately block diagonal. Here, $S^{\dagger}=-S$ is an anti-Hermitian operator. $S$ is constructed perturbatively as:
\begin{equation}
    S=S^{(1)}+S^{(2)}+S^{(3)}+\cdots\ \text{with}\ S^{(m)}\propto \lambda^m
\end{equation}
where $\lambda\sim\mathcal{O}(\lambda_j)$ is a measure of the perturbation strength. The requirement that the transformed Hamiltonian is approximately block-diagonal defines a specific set of equations for $S$. 


Generally, at $m$-th order, the equation determining $S^{(m)}$ is given by:
\begin{equation}
    \left(1-\hat{P}_0\right)\left(V_m+\left[H_0,S^{(m)}\right]\right)\hat{P}_0=0
    \label{Eq:FlowEq_mthOrder}
\end{equation}
where $\hat{P}_0$ is the projector onto the ground-state subspace [Eq.~(\ref{Eq:GSProjector})]; for $m=1$, $V_1=H_1$ and for $m\geq2$,
\begin{equation}
    \begin{split}
        V_m=&\sum_{n_1+\cdots +n_{c_1}=m-1}\frac{1}{c_1!}[\cdots[[H_1,S^{(n_1)}],S^{(n_2)}],\cdots,S^{(n_{c_1})}]\\
        +&\sum_{m_1+\cdots+m_{c_2}=m}\frac{1}{c_2!}[\cdots[[H_0,S^{(m_1)}],S^{(m_2)}],\cdots,S^{(m_{c_2})}]
    \end{split}
    \label{Eq:Vm}
\end{equation}
with $c_1\geq 1$ and $c_2\geq 2$. This equation dictates that the $m$-th order transformed Hamiltonian
\begin{equation}
    H^{(m)}=V_m+\left[H_0,S^{(m)}\right]
    \label{Eq:Hm}
\end{equation}
is block-diagonal. Indeed, a perturbative expansion of the transformed Hamiltonian takes the form of:
\begin{equation}
\begin{split}
    \tilde{H}=&e^{-S}\left(H_0+H_1\right)e^S\\
    =&H_0+H^{(1)}+H^{(2)}+\cdots+H^{(m)}+\cdots
\end{split}
\end{equation}
We aim to look for the solutions of $S^{(m)}$ up to certain order $M$. Then, the transformed Hamiltonian ($\tilde{H}\approx H_0+H^{(1)}+\cdots+H^{(M)}$) is block-diagonal up to $M$-th order.

\subsection{Construction of $S$ for Stabilizer Perturbation}
\label{Sec:AnalyticalConstruct_S}

We show that the anti-Hermitian operator $S$ can be constructed very conveniently in terms of Pauli algebra. We focus on the $m$-th order equation, Eq.~(\ref{Eq:FlowEq_mthOrder}). First, note that $V_m$ in Eq.~(\ref{Eq:Vm}) is determined by $S^{(m^{\prime})}$ with $m^{\prime}<m$. Suppose $\left\{S^{(1)},\cdots,S^{(m-1)}\right\}$ is known already, then $V_m$ can be computed explicitly. At $m$-th order, we employ Eq.~(\ref{Eq:FlowEq_mthOrder}) to solve for $S^{(m)}$. 

It is convenient to parameterize
\begin{equation}
    V_m=\sum_{P_s}c_{P_s}P_s
    \label{Eq:Vm_PauliStringSum}
\end{equation}
as a summation of Pauli strings $P_s\in P_N$, with coefficients $c_{P_s}$. Then, we are able to construct the following particular solution:
\begin{equation}
    S^{(m)}=\sum_{P_s:\ \left[P_s,H_0\right]\neq0}\frac{c_{P_s}}{\Delta E_{P_s}}s_{P_s}P_s
    \label{Eq:S_mthOrder_solution}
\end{equation}
Let us explain several quantities that enter this particular solution: (i) the meaning of the summation; (ii) the Pauli operator $s_{P_s}$; (iii) the energy difference $\Delta E_{P_s}$. 

First of all, the summation in Eq.~(\ref{Eq:S_mthOrder_solution}) is over the Pauli strings $P_s$ present in Eq.~(\ref{Eq:Vm_PauliStringSum}), such that $\left[P_s,H_0\right]\neq0$. Namely, the summation is over the Pauli strings, whose matrix structure contains off-diagonal blocks connecting the unperturbed ground state to some excited states. The Pauli strings that commute with $H_0$ do not enter the construction of $S^{(m)}$. The reason is that those Pauli strings already have the desired block-diagonal structure in the transformed Hamiltonian $H^{(m)}$ [Eq.~(\ref{Eq:Hm})].

Second, for each Pauli string $P_s$ (with $[P_s,H_0]\neq0$), there is a unique excited subspace ``$n$'' such that $\langle n|P_s|0\rangle \neq 0$. Below, we need to determine this excited subspace $n$. Then, we are able to determine the energy difference $\Delta E_{P_s}$ and the Pauli operator $s_{P_s}$.

The excited subspace ``$n$'' such that $\langle n|P_s|0\rangle \neq 0$ can be determined by checking the commutation relation between $P_s$ and all the stabilizer generators. Following the notation in Section~\ref{Sec:StabHam_HilbertSpace}, the excited subspace can be uniquely determined by the indices $n=\{n_i,i=1,2,\cdots,l\}$:
\begin{equation}
    n_i=\left\{\begin{split}
        & 0\ \ \ \text{if}\ \left[P_s,g_i\right]=0\\
        & 1\ \ \ \text{if}\ \left[P_s,g_i\right]\neq 0
    \end{split}\right.
\end{equation}
Given this excited subspace, the energy difference in $S^{(m)}$ [Eq.~(\ref{Eq:S_mthOrder_solution})] can be determined as:
\begin{equation}
    \Delta E_{P_s}=E_n-E_0
\end{equation}
The explicit expression is given by Eq.~(\ref{Eq:EnergyDiff_n&0}). Below, we provide an equivalent expression for this energy difference, see Eq.~(\ref{Eq:EnergyDiff_FlowEq_Sol}) (which is more operational and numerically straightforward).

The last piece of ingredient in $S^{(m)}$ [Eq.~(\ref{Eq:S_mthOrder_solution})] is the Pauli operator $s_{P_s}$. This Pauli operator is picked from the stabilizer Hamiltonian $H_0$ with the requirement that $\left[s_{P_s},P_s\right]\neq 0$. Note that there could be multiple operators that do not commute with $P_s$. It turns out that this choice does not affect the physical observables. Indeed, Eq.~(\ref{Eq:FlowEq_mthOrder}) only uniquely determines the off-diagonal blocks for $S^{(m)}$. In principle, the matrix elements in the diagonal blocks can be arbitrary. We expect that physical observables do not depend on this redundancy in the solution to Eq.~(\ref{Eq:FlowEq_mthOrder}).

\subsubsection{Verification of Solution Eq.~(\ref{Eq:S_mthOrder_solution})}

Now, let us explicitly verify that the construction of $S^{(m)}$ [Eq.~(\ref{Eq:S_mthOrder_solution})] satisfies Eq.~(\ref{Eq:FlowEq_mthOrder}).

First of all, Eq.~(\ref{Eq:FlowEq_mthOrder}) can be written as:
\begin{equation}
    \left(1-\hat{P}_0\right)\left(\sum_{P_s}c_{P_s}P_s+H_0S^{(m)}-S^{(m)}H_0\right)\hat{P}_0=0
\end{equation}
It is enough to consider the following equation:
\begin{equation}
    \left(1-\hat{P}_0\right)\left(P_s+H_0S^{(m)}_{P_s}-S^{(m)}_{P_s}H_0\right)\hat{P}_0=0
    \label{Eq:FlowEq_mthOrder_partial}
\end{equation}
where the solution Eq.~(\ref{Eq:S_mthOrder_solution}) is expressed as:
\begin{equation}
    \left\{\begin{split}
        &S^{(m)}=\sum_{P_s:\ \left[P_s,H_0\right]\neq0}c_{P_s}S^{(m)}_{P_s},\\ &S^{(m)}_{P_s}=\frac{1}{\Delta E_{P_s}}s_{P_s}P_s.
    \end{split}\right.
\end{equation}
Note that if $[P_s,H_0]=0$ implying $\left(1-\hat{P}_0\right)P_s\hat{P}_0=0$, then it is convenient to set the corresponding $S^{(m)}_{P_s}=0$. Hence the summation is over the Pauli strings $P_s$ such that $[P_s,H_0]\neq0$. 

To proceed and to avoid any ambiguity, let us define the following set:
\begin{enumerate}
    \item The set of stabilizer operators/generators that does not commute with $P_s$:
    \begin{equation}
        G_{\langle g_i\rangle}(P_s)=\left\{g\ |\  [g,P_s]\neq 0\ \text{for}\ g\in\{g_1,g_2,\cdots,g_{l}\}\right\}.
    \end{equation}
    This set determines the excited subspace ``$n$'' such that $n_i=1$ if $g_i\in G_{\langle g_i\rangle}(P_s)$ and zero otherwise. The states in this excited subspace satisfy $\langle n|P_s|0\rangle\neq 0$.
    \item The set of Pauli operators $s_i$ that shows up in the stabilizer Hamiltonian $H_0$:
    \begin{equation}
    \mathcal{S}(H_0)=\left\{s_i\in\mathsf{Stab}\ |\ H_0=-\sum_ih_is_i\right\}.
    \label{Eq:Set_H0}
    \end{equation}
    \item The subset of $\mathcal{S}(H_0)$ that does not commute with $P_s$:
    \begin{equation}
    \mathcal{S}(P_s)=\left\{s_i\in \mathcal{S}(H_0)\ |\ \left[P_s, s_i\right]\neq0\right\}.
    \label{Eq:Set_S(Ps)}
    \end{equation}
    This set determines the operator $s_{P_s}$ and the energy difference $\Delta E_{P_s}$. Namely, 
    \begin{equation}
        s_{P_s}\in\mathcal{S}(P_s)
    \end{equation}
    can be any one element of the set $\mathcal{S}(P_s)$. The energy difference can be determined to be
    \begin{equation}
        \Delta E_{P_s}=E_n-E_0=\sum_{i:s_i\in \mathcal{S}(P_s)}2h_i.
        \label{Eq:EnergyDiff_FlowEq_Sol}
    \end{equation}
    Namely, the eigenvalue of $s_i\in \mathcal{S}(P_s)$ is flipped by the Pauli string $P_s$. For each flipped $s_i$, there is an energy difference $2h_i$.
\end{enumerate}

Putting the ingredients above together, Eq.~(\ref{Eq:FlowEq_mthOrder_partial}) is equivalent to:
\begin{equation}
\begin{split}
    0=&\langle n|P_s|0\rangle+\frac{1}{\Delta E_{P_s}}\left(\langle n|H_0s_{P_s}P_s|0\rangle-\langle n|s_{P_s}P_sH_0|0\rangle\right)\\
    =&\langle n|P_s|0\rangle+\frac{E_n-E_0}{\Delta E_{P_s}}\langle n|s_{P_s}P_s|0\rangle
\end{split}
\end{equation}
where $H_0|n\rangle=E_n|n\rangle$ and $H_0|0\rangle=E_0|0\rangle$. Note the following relation:
\begin{equation}
    \langle n|s_{P_s}P_s|0\rangle=-\langle n|P_ss_{P_s}|0\rangle=-\langle n|P_s|0\rangle
\end{equation}
Then, one should be able to verify that Eq.~(\ref{Eq:S_mthOrder_solution}) is one solution to Eq.~(\ref{Eq:FlowEq_mthOrder}).

\subsection{Ground State Expectation Value}
\label{Sec:GS_Exp}

We evaluate the ground state expectation value perturbatively. With the SW transformation above, the unperturbed ground state $|0\rangle$ is approximately the ground state of the transformed Hamiltonian $\tilde{H}$. Equivalently, the \emph{perturbed} ground state $|\text{G.S.}\rangle$ is approximately given by $|\text{G.S.}\rangle\approx e^{S}|0\rangle$. Then, the expectation value of an operator $\hat{O}$ is given by $\langle\text{G.S.}|\hat{O}|\text{G.S.}\rangle\approx\langle 0|e^{-S}\hat{O}e^{S}|0\rangle$, which is evaluated perturbatively:
\begin{equation}
\begin{split}
    \langle\text{G.S.}|\hat{O}|\text{G.S.}\rangle\approx\sum_{m=0}^{M}\langle 0|\hat{O}^{(m)}|0\rangle
\end{split}
\end{equation}
where $M$ is a cutoff in the perturbative expansion; $\hat{O}^{(m)}$ is given by:
\begin{equation}
    \begin{split}
        &\hat{O}^{(0)}=\hat{O}\\
        &\hat{O}^{(m)}=\sum_{m_1+\cdots+m_c=m}\frac{1}{c\ !}[\cdots[[\hat{O},S^{(m_1)}],S^{(m_2)}],\cdots,S^{(m_c)}]
    \end{split}
    \label{Eq:Operator_PertExpansion}
\end{equation}
While formally straightforward, the actual evaluation of the expectation value requires careful analysis on the operators $\hat{O}^{(m)}$.

To evaluate the expectation value perturbatively, we rely on the stabilizer nature of the unperturbed state $|0\rangle$. The strategy is to specify a particular state $|0\rangle$ in the ground-state subspace and express the operator $\hat{O}^{(m)}$ in terms of stabilizer operators. Then, the computation of the expectation value $\langle 0|\hat{O}^{(m)}|0\rangle$ follows directly from the definition of the stabilizer state.

First of all, we consider a specific state in the unperturbed ground-state subspace to avoid any subtlety (especially when dealing with a spontaneous symmetry breaking state). To specify a particular state in the ground-state subspace, $(N-l)$ additional stabilizer oeprators are required $\{g_1,g_2,\cdots,g_l,\tilde{g}_{l+1},\cdots,\tilde{g}_{N}\}$, see Section~\ref{Sec:StabHam_HilbertSpace}.

Second, it is convenient to express the operator $\hat{O}^{(m)}$ as a summation of Pauli strings:
\begin{equation}
    \hat{O}^{(m)}=\sum_{P_s:G_{\langle g_i,\tilde{g}_j\rangle}(P_s)=\emptyset}c_{P_s}P_s+\cdots
    \label{Eq:O_mthOrder_PauliSum}
\end{equation}
Here, we explicitly write down Pauli strings that commute with \emph{all} the stabilizer operators $[P_s,g_i]=0$ and $[P_s,\tilde{g}_j]=0$ for all $i=1,2,\cdots,l$ and $j=l+1,l+2,\cdots,N$. Formally, the summation is over those $P_s$ such that $G_{\langle g_i,\tilde{g}_j\rangle}(P_s)=\emptyset$ is an empty set, where $G_{\langle g_i,\bar g_j\rangle}(P_s)$ is defined as $$G_{\langle g_i,\tilde{g}_j\rangle}(P_s):=\{g\ |\ [g,P_s]\neq0, g\in\{g_1,\cdots,g_l,\tilde{g}_{l+1},\cdots,\tilde{g}_N\}\}.$$ 
The Pauli strings that do not commute with some stabilizer operators are contained in the ``$\cdots$'' in Eq.~(\ref{Eq:O_mthOrder_PauliSum}). Thus, the terms in the dots contribute zero to the expectation value.

Since $P_s$ specified in Eq.~(\ref{Eq:O_mthOrder_PauliSum}) commutes with all stabilizer operators, then $P_s$ is proportional to certain elements in the stabilizer group $\mathsf{Stab}_{\langle g_i,\tilde{g}_j\rangle}=\langle g_1,g_2,\cdots,g_l,\tilde{g}_{l+1},\cdots, \tilde{g}_N\rangle$.
Hence, $P_s$ can be expressed as product of stabilizer operators:
\begin{equation}
    P_s=\text{sign}_{\text{sd}}\left(P_s\right)\prod_{i=1}^lg_i^{n_i}\prod_{j=l+1}^N\tilde{g}_{j}^{n_j},\ \ \ n_{i,j}=0,1
    \label{Eq:PauliString_CB_SB}
\end{equation}
where there is an overall sign, $\text{sign}_{\text{sd}}\left(P_s\right)=\pm$, determined by the multiplication of the stabilizer operators on the right hand side of the equation; the subscript ``sd'' stands for ``stabilizer operator decomposition''.

Since the unperturbed ground state is stabilized by the stabilizer operators $\{g_i,\tilde{g}_j\}$, then by definition: $g_i|0\rangle=|0\rangle$ and $\tilde{g}_j|0\rangle=|0\rangle$
for all $i=1,2,\cdots,l$ and $j=l+1,\cdots,N$. Then, it is straightforward to see the following result: $\langle 0|P_s|0\rangle = \langle 0|\text{sign}_{\text{sd}}\left(P_s\right)\prod_{i=1}^lg_i^{n_i}\prod_{j=l+1}^N\tilde{g}_{j}^{n_j}|0\rangle =\text{sign}_{\text{sd}}\left(P_s\right)$. And correspondingly:
\begin{equation}
    \langle 0|\hat{O}^{(m)}|0\rangle = \sum_{P_s:G_{\langle g_i,\tilde{g}_j\rangle}(P_s)=\emptyset}\text{sign}_{\text{sd}}\left(P_s\right) c_{P_s}
    \label{Eq:Result_Exp_Om}
\end{equation}
Let us emphasize again: $P_s$ in Eq.~(\ref{Eq:O_mthOrder_PauliSum}) and Eq.~(\ref{Eq:Result_Exp_Om}) are Pauli strings that commute with all the stabilizer operators.

Before ending this subsection, we should address the question of how to find the decomposition in Eq.~(\ref{Eq:PauliString_CB_SB}). It turns out this decomposition can be found with the aid of another set of operators: the \emph{de}-stabilizer operators. Numerically, the decomposition can be found efficiently by linear algebra over the field $\mathbb{F}_2$. Then, the sign, $\text{sign}_{\text{sd}}\left(P_s\right)$ in Eq.~(\ref{Eq:PauliString_CB_SB}) and Eq.~(\ref{Eq:Result_Exp_Om}), can be determined after the decomposition is found.

\subsubsection{Finding the decomposition in Eq.~(\ref{Eq:PauliString_CB_SB})}

Focus on Pauli strings $P_s$ that commute with all the stabilizer operators. We discuss how to find the decomposition in Eq.~(\ref{Eq:PauliString_CB_SB}). 

For notational convenience, let us denote the stabilizer operators as $\{g_1,g_2,\cdots,g_l,g_{l+1},\cdots,g_N\}$ (dropping the tilde for the $g_i$'s with $i>l$). Then, the de-stabilizer operators are a set of $N$ operators:
\begin{equation}
    \{dg_1, dg_2,\cdots,dg_l, dg_{l+1},\cdots,dg_N\}
\end{equation}
satisfying the following commutation relation:
\begin{equation}
    \begin{split}
        &\left[dg_i,g_j\right]=0\ \ \ \text{for all}\ i\neq j\\
        &\left[dg_i,g_i\right]\neq0
    \end{split}
    \label{Eq:DeStab_CommutationRelation}
\end{equation}
Namely, in the stabilizer basis, the stabilizer operators can be viewed as ``$Z$'' operators and the de-stabilizer operators are the corresponding ``$X$'' or ``$Y$'' operators. Given the commutation relation above, the de-stabilizer can detect the presence of stabilizer operators in a Pauli string. Indeed, the decomposition in Eq.~(\ref{Eq:PauliString_CB_SB}) can be determined as:
\begin{equation}
    P_s=\text{sign}_{\text{sd}}\left(P_s\right)\prod_{i:[P_s,dg_i]\neq 0}g_i.
\end{equation}
Note that this equation is valid only for Pauli strings that commute with \emph{all} the stabilizer operators.

We should point out that the de-stabilizer operators are generally non-local and thus complicated. Nevertheless, there is a systematic way to construct the de-stabilizer operators, by employing the Smith normal form over $\mathbb{F}_2$. Details are as follows.

First, we need the check matrix $\mathsf{CM}_{\text{gs}}$ for the stabilizer operators. The meaning of this check matrix is that the $i$-th row of $\mathsf{CM}_{\text{gs}}$ corresponds to the Pauli string of $g_i$.

We aim at the check matrix $\mathsf{CM}_{\text{dgs}}$ for the de-stabilizer operators. Namely, the $i$-th row of $\mathsf{CM}_{\text{dgs}}$ corresponds to the Pauli string of $dg_i$. The commutation relation of Eq.~(\ref{Eq:DeStab_CommutationRelation}) translates to an equation for the check matrices:
\begin{equation}
    \mathsf{CM}_{\text{dgs}}\ \hat{L}\ \mathsf{CM}_{\text{gs}}^{\text{T}}\mod2=\mathbb{I}_{N\times N}.
\end{equation}

From now on to the end of this subsection, we work with linear algebra over the field $\mathbb{F}_2$. Namely, all matrix elements are binary valued, taking values of $\{0,1\}$. All arithmetic operations are defined with respect to binary numbers. In particular, the addition is defined modulo 2.

To find $\mathsf{CM}_{\text{dgs}}$, we start with the Smith normal form for the check matrix of stabilizer operators:
\begin{equation}
    P\ \mathsf{CM}_{\text{gs}}\ Q=\begin{bmatrix}
        \mathbb{I}_{N\times N} & 0_{N\times N}
    \end{bmatrix}
    \label{Eq:SmithNF_CM_gs}
\end{equation}
where $P$ is an $N\times N$ invertible binary matrix; $Q$ is a $2N\times 2N$ invertible binary matrix. The right hand side of the equation above follows from the fact $\text{rank}\left(\mathsf{CM}_{\text{gs}}\right)=N$. This Smith normal form decomposition can be found by the method of Gaussian elimination.

Parameterize the matrix $Q$ as $N\times N$ blocks:
\begin{equation}
    Q=\begin{bmatrix}
        Q_{11} & Q_{12}\\
        Q_{21} & Q_{22}
    \end{bmatrix}
\end{equation}
Then, the check matrix for the de-stabilizer operators can be constructed as
\begin{equation}
    \mathsf{CM}_{\text{dgs}}=\begin{bmatrix}
        \left(Q_{21}P\right)^{\text{T}} & \left(Q_{11}P\right)^{\text{T}}
    \end{bmatrix}
    \label{Eq:DeStab_CM_Result}
\end{equation}
The verification of this construction is provided in Appendix~\ref{Sec:Verify_DeStab_CM}. This completes the discussion of perturbative calculation of ground state expectation value.

\subsection{Further Discussions}

Let us emphasize that the stabilizer perturbation calculation introduced above relies on the Pauli algebra and stabilizer nature of the unperturbed state. Hence, the whole process can be implemented numerically in an efficient way with the aid the binary encoding of Pauli strings, see Section~\ref{Sec:NumericalImplementation}. The computational cost is at most polynomial in system size, while exponential in the perturbation order.

One advantage of the binary encoding of Pauli strings is that the matrix/array size scales linearly with the number of qubits. In other words, each element in the check matrix corresponds a specific qubit in the system. This fact buys us convenience when there is specific geometry in the qubit's location, in particular when there is translation invariance. Indeed, one can freely reshape the dimensions of the check matrices in a way that the translation invariance can be implemented quite straightforwardly.

Another important feature is that the whole process is locality preserving. If both $H_0$ and $H_1$ consist of local Pauli operators, then at any finite perturbation order $m\ll L$ (suppose the system size $L$ is large enough), the anti-Hermitian operator $S^{(m)}$ and the effective Hamiltonian $H^{(m)}$ also consist of local terms. The support of each Pauli string in $S^{(m)}$ and $H^{(m)}$ grows at maximum linearly with the perturbation order $m$.

Given the convenience of translation invariance and locality preserving, we can perform the stabilizer perturbation calculation for a large system with hundreds of qubits (on a PC laptop). Note that the unitary $e^{S}$ of local SW transformation is generally non-Clifford. The exact simulation of $e^{S}$ can be extremely hard even with perturbatively constructed $S$ \cite{PhysRevA.70.052328}. Simplification comes from the Taylor expansion of the observables Eq.~(\ref{Eq:Operator_PertExpansion}). The locality preserving nature of the stabilizer perturbation theory further reduces the computational cost.

Lastly, we point out that the whole process of stabilizer perturbation calculation is symmetry preserving. Namely, if the Hamiltonian $H=H_0+H_1$ exhibits certain global symmetry, the anti-Hermitian operator $S$ and the transformed Hamiltonian $\tilde{H}=e^{-S}He^{S}$ are symmetric under the same global symmetry transformation. The reason is that when there is global symmetry, the Pauli strings in $H_{0,1}$ carry zero symmetry charge. Our construction of SW transformation, namely the anti-Hermitian operator $S$, only utilizes the terms in the Hamiltonian $H_{0,1}$. Hence, the anti-Hermitian operator $S$ carries zero symmetry charge and is symmetric. The constructed SW transformation keeps the global symmetry intact.

\section{Local Schrieffer-Wolff Transformation: Numerical Implementation}
\label{Sec:NumericalImplementation}

In this section, we outline the numerical implementation for the stabilizer perturbative calculation. The analytical construction in Section~\ref{Sec:PertStabHam_LocalSWTransformation} gets complicated very soon beyond a few lowest order perturbations. Note that the analytical construction relies on the Pauli algebra. Therefore, an efficient numerical implementation is possible. 

\subsection{Local Schrieffer-Wolff Transformation}

In this subsection, we aim to construct the anti-Hermitian operator $S$ in the local SW transformation. To initiate the computation, the information of the unperturbed stabilizer Hamiltonian $H_0$ as well as the perturbation $H_1$ is required:
\begin{center}
\begin{tabular}{c c|c c c c}
    Input Data & & & Check Matrix &  & Coefficients\\
     $H_0=-\sum_ih_is_i$   &   &  &$\mathsf{CM}_{H_0}$   &  &$\mathsf{Coeff}_{H_0}$\\
     $H_1=\sum_j\lambda_j\mathcal{O}_j$   &   &  &$\mathsf{CM}_{H_1}$   &  &$\mathsf{Coeff}_{H_1}$
\end{tabular}
\end{center}
The anti-Hermitian operator $S$ in the local SW transformation is constructed recursively following the perturbation order $S=\{S^{(1)},S^{(2)},S^{(3)},\cdots\}$.

\subsubsection{The anti-Hermitian operator $S^{(m)}$}
\label{Sec:FirstOrder_sol}

Suppose we already constructed $\{S^{(1)},S^{(2)},\cdots,S^{(m-1)}\}$. To solve for $S^{(m)}$, we follow the steps below:
\newline

\textbf{Zeroth Step:} Compute $V_m$ in Eq.~(\ref{Eq:FlowEq_mthOrder}). At first order, $V_1=H_1$. At higher order $m\geq2$, we use the Pauli algebra to compute $V_m$ according to Eq.~(\ref{Eq:Vm}):
\begin{equation}
    \begin{split}
        V_m=&\sum_{n_1+\cdots +n_{c_1}=m-1}\frac{1}{c_1!}[\cdots[[H_1,S^{(n_1)}],S^{(n_2)}],\cdots,S^{(n_{c_1})}]\\
        +&\sum_{m_1+\cdots+m_{c_2}=m}\frac{1}{c_2!}[\cdots[[H_0,S^{(m_1)}],S^{(m_2)}],\cdots,S^{(m_{c_2})}]
    \end{split}
\end{equation}
with $c_1\geq 1$ and $c_2\geq 2$.
\newline

\textbf{First step:} The central step is to check the commutation relation between $V_m$ and $H_0$:
\begin{equation}
    M_{[V_m,H_0]}=\mathsf{CM}_{V_m}\ \hat{L}\ \mathsf{CM}_{H_0}^{\text{T}}\mod2
\end{equation}
This matrix contains the information for constructing the set of Eq.~(\ref{Eq:Set_S(Ps)}). Without loss of generality, let us focus on one particular row, say $i$-th row corresponding to the $i$-th Pauli string $P_s$ in $V_m$. The set $\mathcal{S}(P_s)$ can be determined to be $\mathcal{S}(P_s)=\{s_j\ |\ M_{[V_m,H_0]}^{ij}=1\}$. Below, we employ the information of $M_{[V_m,H_0]}$ to construct $S^{(m)}$, namely the check matrix $\mathsf{CM}_{S^{(m)}}$ and the coefficients $\mathsf{Coeff}_{S^{(m)}}$.
\newline

\textbf{Second step:} Construct the check-matrix $\mathsf{CM}_{S^{(m)}}$:

First, we need to pick out $s_{P_s}$ in Eq.~(\ref{Eq:S_mthOrder_solution}). This is done by keeping only one nonzero element (if any and either one is fine) in each row of $M_{[V_m,H_0]}$. If certain row is identically zero, then we keep the row unchanged. After this step, we arrive at a reduced matrix:
    \begin{equation}
        M_{[V_m,H_0]}^{\text{red}}=\left. M_{[V_m,H_0]}\right|_{\text{keep 1 nonzero element each row}}
    \end{equation}
Then, the collection of $s_{P_s}$ can be constructed as:
    \begin{equation}
        \mathsf{CM}_{s_{P_s}}=M_{[V_m,H_0]}^{\text{red}}\ \mathsf{CM}_{H_0}
    \end{equation}
In this way, if certain Pauli string commutes with Pauli operators in $[P_s,s_j]=0$ for all $s_j\in \mathcal{S}(H_0)$ [Eq.~(\ref{Eq:Set_H0})], the corresponding $s_{P_s}=I$ is the identity operator. 

The check-matrix $\mathsf{CM}_{S^{(m)}}$ now can be constructed directly:
    \begin{equation}
        \mathsf{CM}_{S^{(m)}}=\left(\mathsf{CM}_{s_{P_s}}+\mathsf{CM}_{V_m}\right)\mod2.
        \label{Eq:FlowEq_Sol_CM}
    \end{equation}
This is to be accompanied with the coefficient vector $\mathsf{Coeff}_{S^{(m)}}$. In particular, certain coefficients will be constructed to be zero, i.e., those associated with the Pauli strings $P_s$ satisfying $[P_s,s_j]=0$ for all $s_j\in \mathcal{S}(H_0)$ [Eq.~(\ref{Eq:Set_H0})].
\newline

\textbf{Third step:} Determine the coefficients $\mathsf{Coeff}_{S^{(m)}}$. Three numerical coefficients needs to be combined: (1) The energy difference $\Delta E_{P_s}$; (2) The phase factor from the multiplication $s_{P_s}P_s$ in Eq.~(\ref{Eq:S_mthOrder_solution}); (3) The coefficients $\mathsf{Coeff}_{V_m}$.

First, the energy difference $\Delta E_{P_s}$ can be computed given the commutation relation $M_{[V_m,H_0]}$. The collection of energy difference $\{\Delta E_{P_s}\}$ is
    \begin{equation}
        \{\Delta E_{P_s}\} = 2\ M_{[V_m,H_0]}\  \text{Abs}\left(\mathsf{Coeff}_{H_0}\right)
    \end{equation}
Just to avoid confusion, ``$\text{Abs}(\mathsf{Coeff}_{H_0})$'' takes the absolute value of each element in $\mathsf{Coeff}_{H_0}$. The result $\{\Delta E_{P_s}\}$ is a column vector, collecting the energy differences associated with each Pauli string in $V_m$.
    
Second, we need to determine the additional phase factor coming from the multiplication $s_{P_s}P_s$, Eq.~(\ref{Eq:S_mthOrder_solution}). 
    
Previously, we determined the check-matrix of the collection of $s_{P_s}$ to be $\mathsf{CM}_{s_{P_s}}$. There are also coefficients associated with the collection of $s_{P_s}$:
    \begin{equation}
        \mathsf{Coeff}_{s_{P_s}}=-M_{[V_m,H_0]}^{\text{red}}\ \text{sign}\left(\mathsf{Coeff}_{H_0}\right)
    \end{equation}
where ``$\text{sign}\left(\mathsf{Coeff}_{H_0}\right)$'' returns a vector consist of signs of each element in  $\mathsf{Coeff}_{H_0}$. The coefficient coming from multiplication of $s_{P_s}P_s$ is given by:
\begin{equation}
        \mathsf{Coeff}_{s_{P_s}P_s}=\mathsf{Coeff}_{s_{P_s}}*\left\{F\left(\mathsf{CM}_{s_{P_s}}^{[i]},\mathsf{CM}_{V_m}^{[i]}\right)\right\}
\end{equation}
where $F\left(\mathsf{CM}_{s_{P_s}}^{[i]},\mathsf{CM}_{V_m}^{[i]}\right)$ is computed for each row $i$ and the result for each row is stacked to form a column vector, just like the coefficient vector. ``$*$'' is the element-wise multiplication.

Lastly, the coefficient $\mathsf{Coeff}_{S^{(m)}}$ now can be computed as:
    \begin{equation}
    \begin{split}
        \mathsf{Coeff}_{S^{(m)}}=&\mathsf{Coeff}_{V_m}\ *\ \mathsf{Coeff}_{s_{P_s}P_s} \\
        *&\ \{\Delta E_{P_s}\neq 0\} / \{\Delta E_{P_s}+\eta\}
    \end{split}
        \label{Eq:FlowEq_Sol_Coeff}
    \end{equation}
where $\eta=0^+$ is a small number to avoid numerical division by zero; ``$*$'' is element-wise multiplication and ``$/$'' the the element-wise division. The factor $\{\Delta E_{P_s}\neq 0\}$ in the numerator sets to zero the coefficients associated with the Pauli strings $P_s$ satisfying $[P_s,s_j]=0$ for all $s_j\in \mathcal{S}(H_0)$ [Eq.~(\ref{Eq:Set_H0})].

To this end, we are able to obtain the $m$-th order solution $S^{(m)}$, whose check-matrix is given by Eq.~(\ref{Eq:FlowEq_Sol_CM}) and coefficients given by Eq.~(\ref{Eq:FlowEq_Sol_Coeff}).

\subsection{Ground State Expectation Values}

Now we are ready to compute the ground state expectation values of an operator $\hat{O}$.

We first need to specify a particular ground state; then, we need to calculate the perturbative expansion of $e^{-S}\hat{O}e^{S}$; lastly, we evaluate the expectation value order by order.

\subsubsection{Ground State}

As mentioned, to compute the ground state expectation, we need to pick a particular state from the ground-state subspace. This is done by finding a few additional stabilizer operators $\{\tilde{g}_{l+1},\tilde{g}_{l+2},\cdots,\tilde{g}_{N}\}$. 

In many cases, those additional operators can be constructed based on physical knowledge of the system under consideration. In the end, we obtain a full set of stabilizer operators $\{g_1,g_2,\cdots,g_l,\tilde{g}_{l+1},\cdots,\tilde{g}_{N}\}$, represented as check-matrix and coefficients. As discussed in Section~\ref{Sec:GS_Exp}, the information of de-stabilizer operators can be very useful. Hence, we need the following data to initiate the calculation of ground state expectation values:
\begin{center}
    \begin{tabular}{c c c}
    \hline
        Stabilizer Operators &  Check Matrix & \ Coeff.\\
         $\{g_1,g_2,\cdots,g_l,\tilde{g}_{l+1},\cdots,\tilde{g}_{N}\}$ & $\mathsf{CM}_{\text{gs}}$ & \ $\mathsf{Coeff}_{\text{gs}}$\\
          & & \\
         de-Stabilizer Operators &   & \\
         $\{dg_1,dg_2,\cdots,dg_l,dg_{l+1},\cdots,dg_{N}\}$ & $\mathsf{CM}_{\text{dgs}}$ & N.A.\\
          & & \\
         \hline
    \end{tabular}
\end{center}
where the coefficients for de-stabilizer operators are not important. Note that the de-stabilizer operator is not an independent piece of information. As discussed in Section~\ref{Sec:GS_Exp} and Appendix~\ref{Sec:Verify_DeStab_CM}, $\mathsf{CM}_{\text{dgs}}$ can be constructed from $\mathsf{CM}_{\text{gs}}$.

In the most general cases, the additional stabilizer operators $\{\tilde{g}_{l+1},\tilde{g}_{l+2},\cdots,\tilde{g}_{N}\}$ may not be easy to construct heuristically. Through binary linear algebra, one can find the centralizer $\mathsf{Central}\left(g_1,\cdots,g_{l}\right)$, consist of operators that commute with all $g_i$ with $i=1,2,\cdots,l$. Then, one has to investigate $\mathsf{Central}\left(g_1,\cdots,g_{l}\right)$ to look for the desired additional stabilizer operators $\{\tilde{g}_{l+1},\tilde{g}_{l+2},\cdots,\tilde{g}_{N}\}$. 

\subsubsection{Ground State Expectation Value}

We evaluate the ground state expectation value order by order. Below, we demonstrate the evaluation of $m$-th order correction $\langle 0|\hat{O}^{(m)}|0\rangle$.
\newline

\textbf{Zeroth step:} We employ the Pauli algebra to compute the perturbative expansion of $e^{-S}\hat{O}e^{S}$:
\begin{equation}
    \begin{split}
        &\hat{O}^{(0)}=\hat{O}\\
        &\hat{O}^{(m)}=\sum_{m_1+\cdots+m_c=m}\frac{1}{c\ !}[\cdots[[\hat{O},S^{(m_1)}],S^{(m_2)}],\cdots,S^{(m_c)}]
    \end{split}
\end{equation}

\textbf{First step:} We write the operator as summation of Pauli strings:
\begin{equation}
    \hat{O}^{(m)}=\sum_{P_s}c_{P_s}P_s+\cdots
\end{equation}
and aim to extract the Pauli strings that commutes with \emph{all} the stabilizer operators (those written out explicitly in the equation above). This is done by directly checking the commutation relation between each Pauli string in $\hat{O}^{(m)}$ and the stabilizer operators:
\begin{equation}
    M_{[\hat{O}^{(m)},g]}=\hat{O}^{(m)}\ \hat{L}\ \mathsf{CM}_{\text{gs}}^{\text{T}}\mod2.
\end{equation}
When certain row of $M_{[\hat{O}^{(m)},g]}$ is all zero, then the corresponding Pauli string commutes with \emph{all} the stabilizer operators. Below, we perform further operations on those Pauli strings:
\begin{equation}
    \hat{O}^{(m)}_c=\sum_{P_s}c_{P_s}P_s
\end{equation}
whose check matrices and coefficients is collectively denoted as $\mathsf{CM}_{\hat{O}^{(m)}_c}$ and $\mathsf{Coeff}_{\hat{O}^{(m)}_c}$.
The subscript ``$c$'' (as in $\hat{O}^{(m)}_c$) dictates that we keep only the Pauli strings that commutes with \emph{all} the stabilizer operators. 
\newline

\textbf{Second step:} The Pauli strings kept in $\hat{O}^{(m)}_c$ can be written as products of stabilizer operators. Now, we aim to find the additional sign in such decomposition. This can be done quite straightforwardly by computing the commutation relation between $\hat{O}^{(m)}_c$ and the \emph{de}-stabilizer operators:
\begin{equation}
    M_{\text{sd}}=\mathsf{CM}_{\hat{O}^{(m)}_c}\ \hat{L}\ \mathsf{CM}_{\text{dgs}}^{\text{T}}\mod 2
\end{equation}
where the subscript ``sd'' stands for ``stabilizer operator decomposition''. The meaning of the matrix $M_{\text{sd}}$ is as follows. Suppose we focus on certain Pauli string $P_s$ corresponding to $i$-th row of $\mathsf{CM}_{\hat{O}^{(m)}_c}^{[i]}$. Then, this Pauli string can be written as:
\begin{equation}
    P_s=\text{sign}_{\text{sd}}\left(P_s\right)\prod_{j:\ M_{\text{sd}}^{ij}=1}g_j
\end{equation}
where the product is over the stabilizer operators $g_j$ when the matrix element is identity $M_{\text{sd}}^{ij}=1$. In addition, there is an overall sign, 
\begin{equation}
    \text{sign}_{\text{sd}}\left(P_s\right)=\pm
\end{equation}
which can be computed by explicitly performing the multiplication $\prod_{j:\ M_{\text{sd}}^{ij}=1}g_j$ via Pauli algebra. This process can be done for each row in $\mathsf{CM}_{\hat{O}^{(m)}_c}$. The additional sign coming from stabilizer decomposition is collectively denoted as $\mathsf{Sign}_{\text{sd}}\left(\mathsf{CM}_{\hat{O}^{(m)}_c}\right)$.

\textbf{Third step:} Given the stabilizer operator decomposition above, it is straightforward to evaluate the expectation value:
\begin{equation}
\begin{split}
    \langle 0|\hat{O}^{(m)}|0\rangle=&\langle 0|\hat{O}^{(m)}_c|0\rangle
    =\sum_{P_s}c_{P_s}\text{sign}_{\text{sd}}\left(P_s\right)
\end{split}
\end{equation}
where we used $\langle0|\prod_j g_j|0\rangle=1$ for any product of stabilizer operators. More numerical friendly equation would be:
\begin{equation}
    \langle 0|\hat{O}^{(m)}|0\rangle=\text{Sum}\left[\mathsf{Coeff}_{\hat{O}^{(m)}_c}*\mathsf{Sign}_{\text{sd}}\left(\mathsf{CM}_{\hat{O}^{(m)}_c}\right)\right]
\end{equation}
This concludes the process of perturbatively computing ground state expectation value of certain operator $\hat{O}$.

\section{Application I: Transverse Field Ising Chain}
\label{Sec:Application:TFIC}

The first application is the ferromagnetic transverse field Ising chain in one dimension:
\begin{equation}
    H=-\sum_j Z_jZ_{j+1}-h\sum_{j}X_j
\end{equation}
For demonstration purpose, we start from the ferromagnetic (FM) phase with small transverse field $|h|<1$. The unperturbed ground-state subspace is stabilized by the $(N-1)$ Ising couplings $\{Z_1Z_2,Z_2Z_3,\cdots,Z_{N-1}Z_N\}$ for a chain with $N$ qubits. Hence, the ground-state subspace is two-fold degenerate (as is well-known that the Ising FM phase exhibits spontaneous $\mathbb{Z}_2$ symmetry breaking).

To further clarify the stabilizer perturbation calculation, we first perform analytical calculation to second order. Then, we perform the numerical perturbative calculation on a finite chain of $N=100$ qubits with periodic boundary condition (PBC) to $10$-th order. We compare the perturbation result with DMRG, which is done with iDMRG algorithm with the aid of TeNPy package \cite{tenpy2024}.

\subsection{Analytical Calculation}

Now, we demonstrate the process of stabilizer perturbation calculation with analytical calculation to second order.

First step is to determine $S^{(1)}$. The equation determining $S^{(1)}$ reads:
\begin{equation}
    \left(1-\hat{P}_0\right)\left(-h\sum_{j}X_j+\left[-\sum_j Z_jZ_{j+1},S^{(1)}\right]\right)\hat{P}_0=0
\end{equation}
Note that each $X_j$ anti-commute with $Z_{j-1}Z_j$ and $Z_jZ_{j+1}$. Hence, we can construct the following quantity:
\begin{equation}
    S^{(1)}_j=-\frac{h}{4}Z_{j-1}Z_jX_j=-i\frac{h}{4}Z_{j-1}Y_j
\end{equation}
The full solution is given by $S^{(1)}=\sum_jS^{(1)}_j$.

The second step is to determine $S^{(2)}$. The equation determining $S^{(2)}$ reads:
\begin{equation}
    \begin{split}
        &\left(1-\hat{P}_0\right)\left(V_2+\left[-\sum_j Z_jZ_{j+1},S^{(2)}\right]\right)\hat{P}_0=0\\
        & V_2=\left[-h\sum_{j}X_j, S^{(1)}\right]+\frac{1}{2}\left[\left[-\sum_j Z_jZ_{j+1}, S^{(1)}\right], S^{(1)}\right]
    \end{split}
\end{equation}
Direct calculation shows that
\begin{equation}
    V_2=\sum_j\left(-\frac{1}{4}h^2Z_jZ_{j+1}+\frac{3}{8}h^2Y_jY_{j+1}-\frac{1}{8}h^2Z_{j-1}X_{j}X_{j+1}Z_{j+2}\right)
\end{equation}
with three typical Pauli strings: (1) $Z_jZ_{j+1}$ is block diagonal already; (2) $Y_jY_{j+1}$ and $Z_{j-1}X_{j}X_{j+1}Z_{j+2}$ both anti-commute with two Ising couplings $\{Z_{j-1}Z_{j},Z_{j+1}Z_{j+2}\}$. Then, we can construct $S^{(2)}$ as:
\begin{equation}
    \begin{split}
        S^{(2)}=&\sum_j\frac{1}{4}Z_{j-1}Z_j\left(\frac{3}{8}h^2Y_jY_{j+1}-\frac{1}{8}h^2Z_{j-1}X_jX_{j+1}Z_{j+2}\right)\\
        =&\sum_j\left(-i\frac{3}{32}h^2Z_{j-1}X_jY_{j+1}-i\frac{1}{32}h^2Y_jX_{j+1}Z_{j+2}\right)
    \end{split}
\end{equation}

To this end, we can compute the transformed Hamiltonian to second order:
\begin{equation}
\begin{split}
    \tilde{H}\approx &\sum_j\left[-\left(1+h^2/4\right)Z_jZ_{j+1}-\frac{1}{2}hX_j+\frac{1}{2}hZ_{j-1}X_jZ_{j+1}\right]\\
    +&\sum_j\frac{1}{8}h^2\left[Y_jY_{j+1}+Z_{j-1}X_jX_{j+1}Z_{j+2}\right]+\mathcal{O}(h^3)
\end{split}
\end{equation}
which can be verified to be block-diagonal up to second order in the transverse field $h$.

A few features to notice. First, the global $\mathbb{Z}_2$ symmetry is maintained in the whole process of calculation. Indeed, the only the terms in $H_0$ and $H_1$, which are symmetric, are employed in the calculation. Hence, we expect the symmetry can be maintained to all orders of perturbation. Second, the Pauli strings in $S$ and $\tilde{H}$ remain local. Nonetheless, the size of the Pauli strings grows with perturbation order. At the second order, four qubit terms are generated. Lastly, the translation invariance allows us to keep track of a few terms at each step of the calculation. Hence, the calculation for large system is possible.

To complete the demonstration, let us compute the magnetization $\langle \text{G.S.}|Z_j|\text{G.S.}\rangle$. First step is to compute the perturbative expansion of $\tilde{Z}_j=e^{-S}Z_je^{S}$. The result up to second order is given by:
\begin{equation}
    \begin{split}
        &\tilde{Z}^{(0)}_j=Z_j, \ \ \ \tilde{Z}^{(1)}_j=-\frac{1}{2}hZ_{j-1}X_j\\
        &\tilde{Z}^{(2)}_j=-\frac{1}{8}h^2Z_j-\frac{1}{16}Z_{j-2}X_{j-1}X_j+\frac{5}{16}h^2Z_{j-1}Y_jY_{j+1}\\
        &\ \ \ \ \ \ \ \ \ \ \ +\frac{1}{16}h^2Z_{j-2}X_{j-1}Y_jZ_{j+1}+\frac{1}{16}h^2Z_{j-1}Y_jX_{j+1}Z_{j+2}
    \end{split}
\end{equation}
To obtain expectation value, we need to specify the ground state. There are a few illustrative options:
\begin{enumerate}
    \item All spin-up state $|\Uparrow\rangle=|\uparrow\uparrow\cdots\uparrow\rangle$: this state is stabilized by 
    \begin{equation}
    \begin{split}
        &\{g_1,g_2,\cdots,g_{N-1},g_N\}\\
        =&\{Z_1Z_2,Z_2Z_3,\cdots,Z_{N-1}Z_N,Z_N\}
    \end{split}
    \end{equation}
    Only $Z_j$ terms in $\tilde{Z}_j$ commute with all the stabilizer operators. The stabilizer operator decomposition is $Z_j=\prod_{i=j}^Ng_i$. One can check this decomposition by noticing that the de-stabilizer operators are $dg_j=\prod_{i=1}^jX_i$. The expectation value can be evaluated as:
    \begin{equation}
    \begin{split}
        \langle \Uparrow|\tilde{Z}_j|\Uparrow\rangle\approx &\left(1-\frac{1}{8}h^2\right)\langle \Uparrow|\prod_{i=j}^Ng_i|\Uparrow\rangle\\
        =&1-\frac{1}{8}h^2.
        \label{Eq:TFIC_Magnetization}
    \end{split}
    \end{equation}
    Hence, magnetization decreases with increasing transverse field as expected.

    \item All spin-down state $|\Downarrow\rangle=|\downarrow\downarrow\cdots\downarrow\rangle$: this state is stabilized by 
    \begin{equation}
    \begin{split}
        &\{g_1,g_2,\cdots,g_{N-1},g_N\}\\
        =&\{Z_1Z_2,Z_2Z_3,\cdots,Z_{N-1}Z_N,-Z_N\}
    \end{split}
    \end{equation}
    Notice the minus sign in the last stabilizer operator. Again, only $Z_j$ terms commute with all the stabilizer operator. Now, the stabilizer operator decomposition is given by $Z_j=-\prod_{i=j}^Ng_i$ (note the minus sign). The expectation value can be evaluated as:
    \begin{equation}
    \begin{split}
        \langle \Downarrow|\tilde{Z}_j|\Downarrow\rangle\approx &\left(1-\frac{1}{8}h^2\right)\langle \Downarrow|-\prod_{i=j}^Ng_i|\Downarrow\rangle\\
        =&-1+\frac{1}{8}h^2.
    \end{split}
    \end{equation}
    This result is opposite to the result for all spin up state as should be;

    \item Greenberger–Horne–Zeilinger (GHZ) state $|\text{GHZ}\rangle=\frac{1}{\sqrt{2}}\left(|\Uparrow\rangle+|\Downarrow\rangle\right)$: this state is stabilized by:
    \begin{equation}
        \begin{split}
            &g_i=Z_iZ_{i+1},\ \ \ \  i=1,2,\cdots,N-1\\
            &g_N=X_1X_2\cdots X_N=\prod_{i=1}^NX_i
        \end{split}
    \end{equation}
    As one can check explicitly, all terms in $\tilde{Z}_j$ anti-commute with $g_N$. Therefore, the expectation value vanishes $\langle \text{GHZ}|\tilde{Z}_j|\text{GHZ}\rangle=0$. There is a symmetry reason behind. Namely, $g_N$ is the global $\mathbb{Z}_2$ symmetry charge/transformation. The operator $S$ by construction is symmetric. Meanwhile, $Z_j$ as well as $\tilde{Z}_j$ are not symmetric operator, i.e., carrying nonzero symmetry charge. Hence, their expectation value with respect to the symmetric GHZ state, which carries a fixed symmetry charge, is identically zero.
\end{enumerate}

As one can see that the perturbation calculation can get complicated at high orders. Hence, below we resort to numerical calculations.

\subsection{Numerical Results}

\begin{figure}[t]    
    \centering
    \includegraphics[width = 0.9\linewidth]{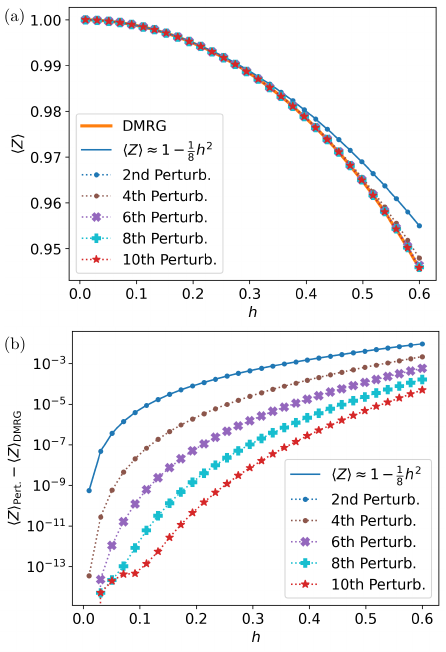}
    \caption{\textbf{Transverse field Ising chain:} (a) Magnetization $\langle Z\rangle$ vs. transverse field strength $h$; (b) Difference between perturbation result and DMRG result.}
    \label{fig:FM_TFIC_Magnetization}
\end{figure}

\begin{figure*}
    \centering
    \includegraphics[width=\textwidth]{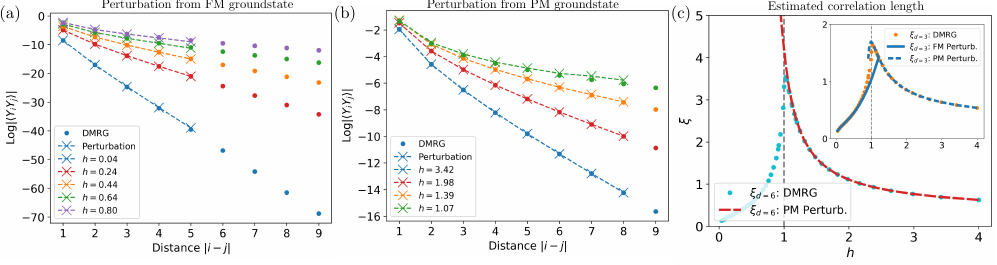}
    \caption{\textbf{Transverse field Ising chain} (a-b) Static correlation funtion $\text{Log}|\langle Y_iY_j\rangle|$ perturbatively computed from (a) ferromagnetic (FM) ground state $|\uparrow\uparrow\cdots\uparrow\rangle$ and (b) paramagnetic (PM) ground state $|\rightarrow\rightarrow\dots\rightarrow\rangle$. Dash-cross lines are the perturbation result; dots are the DMRG data. (c) Estimated correlation length $\xi_d=\left(\text{Log}|\langle Y_0Y_{d}\rangle|-\text{Log}|\langle Y_0Y_{d+1}\rangle|\right)^{-1}$ [Eq.~(\ref{Eq:CorrLength_estimate}]. $\xi_{d=6}$ (main panel) and $\xi_{d=3}$ (inset) are calculated. Both results exhibit a tendency of diverging correlation length upon approaching the critical point $h_c=1$.}
    \label{fig:TFIC_YYCorr_CorrLength}
\end{figure*}

Among the ground-state subspace, we choose the all spin-up state as the unperturbed ground state:
\begin{equation}
\begin{split}
    &|0\rangle=|\uparrow\uparrow\cdots\uparrow\rangle \\
    &\{g_1,g_2,\cdots,g_N\}=\{Z_1Z_2,Z_2Z_3,\cdots,Z_{N-1}Z_N,Z_N\}
\end{split}
\end{equation}
which is a state stabilized by the operators listed in the second line above. In the current example, it is also quite straightforward to find the de-stabilizer operators:
\begin{equation}
    \{dg_1,dg_2,\cdots,dg_N\}:\ dg_j=\prod_{i=1}^jX_i
\end{equation}
To this end, we have all the ingredients to perform the perturbative calculation.

One question to ask is how magnetization changes upon increasing the strength of the transverse field $h$. Namely, we aim to compute the expectation value of $Z$ operator at some site on the chain. In Figure.~\ref{fig:FM_TFIC_Magnetization}, we computed the magnetization up to 10-th order perturbation. Note that the second order result indeed is given by the analytical result in Eq.~(\ref{Eq:TFIC_Magnetization}). Also note that the contribution to magnetization from odd order of perturbation vanishes. Hence only the results up to even order is plotted. 

In Figure~\ref{fig:FM_TFIC_Magnetization}(a), one can clearly observe that starting from 4th order, the perturbation result is hardly distinguishable from DMRG result. Figure~\ref{fig:FM_TFIC_Magnetization}(b) shows that the difference between perturbation and DMRG results decreases with the perturbation order. For moderate transverse field $h\lesssim 0.5$, the deviation between 10th order perturbation and DMRG can be smaller than $\sim10^{-6}$.

Moreover, static correlation function can be computed in exactly the same way. As an example, we compute the static $\langle Y_iY_j\rangle$ correlation function, Figure~\ref{fig:TFIC_YYCorr_CorrLength}. As shown in Figure~\ref{fig:TFIC_YYCorr_CorrLength}(a-b), the correlation function can be computed perturbatively from both the ferromagnetic (FM) ground state $|\uparrow\uparrow\cdots\uparrow\rangle$ and the paramagnetic (PM) ground state $|\rightarrow\rightarrow\dots\rightarrow\rangle$. The perturbative results exhibit good match with the DMRG calculation. The short-coming of the perturbative calculation is that the correlation function can only be calculated for small distance $|i-j|$. The reason behind is related to the perturbative expansion of local operators under local SW transformation, $\tilde{O}=e^{-S}\hat{O}e^S$. Indeed, the perturbative expansion of a local operator $\tilde{O}$ in Eq.~(\ref{Eq:Operator_PertExpansion}) has finite support. As a consequence, the static correlation function reduces to $\langle \tilde{O}_1\tilde{O}_2\rangle=\langle\tilde{O}_1\rangle\langle\tilde{O}_2\rangle$ when the separation between two operators is much larger than their support, $d_{(\tilde{O}_1,\tilde{O}_2)}\gg \text{Supp}(\tilde{O}_{1,2})$.

Nonetheless, useful information, such as correlation length, can be estimated from the correlation function, Figure~\ref{fig:TFIC_YYCorr_CorrLength}(c). We estimate the correlation length using the correlation function at finite distance:
\begin{equation}
    \xi_d=\left(\text{Log}|\langle Y_0Y_{d}\rangle|-\text{Log}|\langle Y_0Y_{d+1}\rangle|\right)^{-1}
    \label{Eq:CorrLength_estimate}
\end{equation}
For instance, we can employ the data of correlation function at distances $d=3$ and $d=6$ in Figure~\ref{fig:TFIC_YYCorr_CorrLength}(a-b) separately. The result of $\xi_{d=3,6}$ are shown in Figure~\ref{fig:TFIC_YYCorr_CorrLength}(c). Away from the critical field strength $h_c=1$, the estimated correlation length is quite consistent between perturbative calculation and DMRG result. Close to the the critical point $h_c=1$, the perturbative result shows a tendency of diverging correlation length, as expected. The diverging correlation length upon approaching the critical point is most transparent from the paramagnetic perturbative calculation.

To conclude this section, we performed the perturbative calculation for transverse field Ising chain. The perturbative calculation shows good agreement with DMRG result. The stabilizer perturbation calculation is not limited to one-dimensional system. Below, we apply the stabilizer perturbation theory to toric code in two dimensions on square lattice and kagome lattice.

\section{Application II: 2D Toric Code on Square Lattice}
\label{Sec:Application_TC_squarelattice}

\begin{figure}[t]
    \centering
    \includegraphics[width=\linewidth]{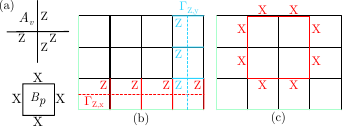}
    \caption{\textbf{Toric code on square lattice} (a) The vertex term $A_v$ and plaquette term $B_p$, defining the toric code Hamiltonian; (b) Two global $Z$ loops; (c) An example of $X_{\text{loop}}$ operator, consist of a $2\times 2$ tiling of the plaquette term $B_p$ with a circumference of 8 qubits.}
    \label{fig:TC_AvBp_GlobalLoops}
\end{figure}

\begin{figure*}
    \centering
    \includegraphics[width=\linewidth]{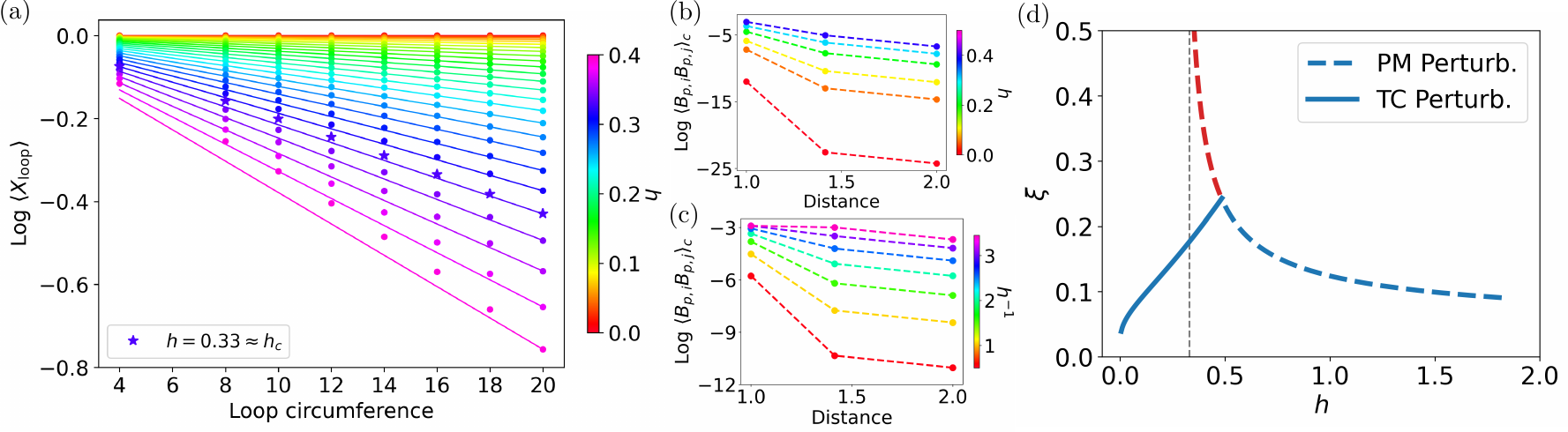}
    \caption{\textbf{Toric code coupled to Zeeman field} (a) Dots: fourth order perturbative data point of $\text{Log}\langle X_{\text{loop}}\rangle$ vs. loop circumference at various Zeeman field strength $h$. Loop circumference is defined as the number of qubits in the loop operator. To aid visualization, straight lines connecting origin and the last data point were drawn. (b-c) Correlation function between plaquette operators $\text{Log}\langle B_{p,i}B_{p,j}\rangle$ computed from (b) toric code (TC) ground state at weak field and (c) paramagnetic ground state at strong field. (d) Correlation length estimated from the first two data points in (b-c).}
    \label{fig:TC_PL_BBCorr_CorrelationLength}
\end{figure*}

The second example is the toric code in two dimensions on a square lattice \cite{Kitaev1997QECwImperfectGates,kitaev1997quantum,KITAEV20032}. The toric code Hamiltonian is given by:
\begin{equation}
    H_0=-\sum_vA_v-\sum_pB_p
\end{equation}
where the vertex terms are $A_v=\prod_{i\in v}Z_i$ and the plaquette terms are $B_p=\prod_{i\in p}X_i$, see Figure~\ref{fig:TC_AvBp_GlobalLoops}(a). Below, we consider two scenarios: (i) Toric code coupled to Zeeman field; (ii) Toric code bilayer with inter-layer Ising coupling. In both scenarios, the perturbation can drive a condensation of certain anyon \cite{annurev:/content/journals/10.1146/annurev-conmatphys-033117-054154}. We use loop operators to probe the confinement transition associated with the anyon condensation.

\subsection{Zeeman field perturbation}

Consider coupling the single layer toric code to Zeeman field:
\begin{equation}
    H_1=-h\sum_i Z_i
    \label{Eq:TC_ZeemanPerturbation}
\end{equation}
It is well-known that when $|h|>h_c$, the Zeeman field can induce the anyon condensation and drive the system into a paramagnetic phase. The anyon condensation can be probed by a confinement transition in the loop operators. We carry out the perturbative calculation to probe such transition.

The perturbative calculation is done for a system of $10\times 10$ unit cells, namely 200 qubits, with PBC. We choose the unperturbed ground state to be stabilized by the two global $Z$-loops, namely $\Gamma_{Z,x/y}$ [Figure~\ref{fig:TC_AvBp_GlobalLoops}(b)], in addition to the independent vertex and plaquette terms. Indeed, the two global $Z$-loops, $\Gamma_{Z,x/y}$, commute with the perturbation Eq.~(\ref{Eq:TC_ZeemanPerturbation}), and thus are good quantum numbers for the perturbed toric code.

The expectation values of ``local'' $X_{\text{loop}}$ operators are calculated. Figure~\ref{fig:TC_AvBp_GlobalLoops}(c) is an example of $X_{\text{loop}}$ operator, consist of a $2\times 2$ tiling of the plaquette term $B_p$ with a circumference of 8 qubits. We computed $X_{\text{loop}}$ operators ranging from a single plaquette to a $5\times 5$ tiling of plaquette operators.

The result is summarized in Figure~\ref{fig:TC_PL_BBCorr_CorrelationLength}. In Figure~\ref{fig:TC_PL_BBCorr_CorrelationLength}(a), the dots are the data point generated from fourth order perturbation calculation. We add straight lines connecting origin to the last data points to aid visualization.

The critical field strength was shown to be around $h_c\approx 0.33$ in previous studies \cite{He1990HTClusterIsing,PhysRevLett.106.107203,zeng2019quantum}. In the current perturbative calculation, we find that when the field strength is small $h< h_c$, the expectation value of $X_{\text{loop}}$ operator follows a perimeter law \cite{annurev:/content/journals/10.1146/annurev-conmatphys-040721-021029},
\begin{equation}
    \langle X_{\text{loop}}\rangle\sim \text{Exp}\left(-\alpha L\right)\ \ \ \ \text{when}\ h<h_c,
\end{equation} 
where $\alpha$ is some coefficient and $L$ is the circumference of the loop. In the calculation of Figure~\ref{fig:TC_PL_BBCorr_CorrelationLength}(a), we defined the circumference as the number of qubits on the loop. A deviation from the perimeter law is observed when the Zeeman field strength is large $h\gtrsim h_c$. Indeed, the perturbative calculation is able to capture the confinement transition in the loop operators and provide good estimation of the critical point.

The confinement transition is accompanied with a diverging correlation length. The connected part of the plaquette correlation function $\langle B_{p,i}B_{p,j}\rangle_c=\langle B_{p,i}B_{p,j}\rangle-\langle B_{p,i}\rangle\langle B_{p,j}\rangle$ is computed from toric code ground state [Figure~\ref{fig:TC_PL_BBCorr_CorrelationLength}(b)] and paramagnetic ground state [Figure~\ref{fig:TC_PL_BBCorr_CorrelationLength}(c)] separately. The correlation length is estimated as shown in Figure~\ref{fig:TC_PL_BBCorr_CorrelationLength}(d). Indeed, the correlation length increases away from the zero field and infinite field limit. More interestingly, in the calculation of perturbed paramagnetic state, the correlation length exhibits a tendency towards infinity upon approaching the critical field strength $h_c\approx 0.33$.

\subsection{Toric code bilayer}

\begin{figure}[t]
    \centering
    \includegraphics[width=\linewidth]{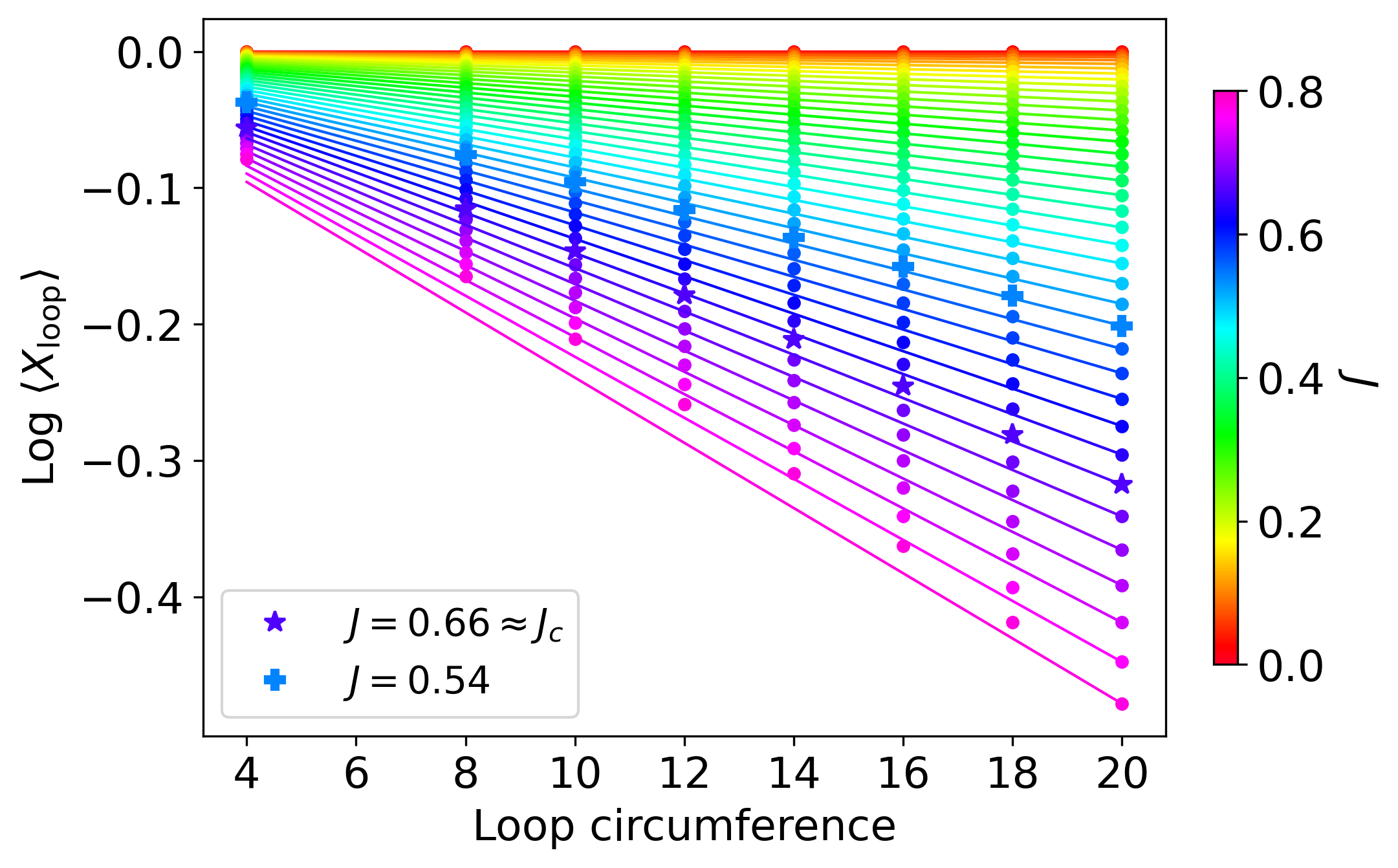}
    \caption{\textbf{Toric code bilayer with inter-layer Ising coupling} Dots: second order perturbative data point of $\text{Log}\langle X_{\text{loop}}\rangle$ vs. loop circumference at various Ising coupling strength $J$. Loop circumference is defined as the number of qubits in the loop operator. To aid visualization, straight lines connecting origin and the last data point were drawn.}
    \label{fig:TCB_Ising_Xloop}
\end{figure}

We can also consider two layers of toric code, coupled by inter-layer Ising coupling:
\begin{equation}
    H_1=-\sum_iZ_i^{(1)}Z_i^{(2)}
\end{equation}
where the superscript labels the two layers. Such interlayer coupling can also induce anyon condensation and drive a quantum phase transition from two copies of toric code to a single toric code ground state \cite{PhysRevB.100.235115,PhysRevB.102.214422}. The $X_{\text{loop}}$ operator in Figure~\ref{fig:TC_AvBp_GlobalLoops}(c) defined in either layer turns out to be a good probe for such transition.

We carried out second order perturbation for a system of $10\times 10$ unit cells, namely 400 qubits, with PBC. The result is summarized in Figure~\ref{fig:TCB_Ising_Xloop}. It is quite clear that when the inter-layer coupling is weak $J\lesssim 0.54$, the expectation value of $X_{\text{loop}}$ exhibits a perimeter law. Deviation from perimeter law can be observed for larger coupling strength $J\gtrsim 0.54$.

Note that based on previous studies, the critical coupling strength is around $J_c\approx 0.66$ \cite{PhysRevB.102.214422}. The current estimation of critical point $J_*\approx 0.54$ is reasonably close given that this is only \emph{second} order perturbation result.

\section{Application III: 2D Toric Code on Kagome Lattice}
\label{Sec:Application_TC_kagome}
Motivated by the relevance of the kagome lattice to spin liquid candidate materials and to demonstrate the applicability of our approach in more complex geometries, we consider models defined on the kagome lattice as our last example. More specifically, we study toric code on kagome lattice perturbed by two types of perturbations separately, including nearest-neighbor (NN) Ising coupling and NN Heisenberg coupling. In all cases, a tendency toward confinement for certain types of anyons is suggested by the perturbation calculations.

\begin{figure}
    \centering
    \includegraphics[width=\linewidth]{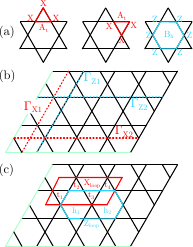}
    \caption{\textbf{Toric code on kagome lattice} (a) Triangle operators $A_t=\prod_{i\in t}X$ is the product of $X$ operators on the vertices of each triangle. Hexagon operators $B_p=\prod_{i\in h}Z$ is the product of $Z$ operators on the vertices of each hexagon. (b) Four types of global loops. Either global $X$-loops $(\Gamma_{X1},\Gamma_{X2})$ or global $Z$-loops $(\Gamma_{Z1},\Gamma_{Z2})$ (not all four loops together) is need to specify a specific ground state. (c) Examples of local loop operators, where $X_{\text{loop}}=A_{t_1}A_{t_2}A_{t_3}A_{t_4}$ and $Z_{\text{loop}}=B_{h_1}B_{h_2}$.}
    \label{fig:KagomeTC_AtBh_GLoops_LLoops}
\end{figure}

On a kagome lattice, toric code Hamiltonian can be defined as:
\begin{equation}
    H_0=-\sum_{t}A_t-\sum_{h}B_h
    \label{Eq:KTC}
\end{equation}
where qubits locate at the vertices/sites of the lattice (rather than the links) ; $\sum_{t,h}$ is the summation over all triangles (``$t$'') and hexagons (``$h$''). As shown in Figure~\ref{fig:KagomeTC_AtBh_GLoops_LLoops}(a), the triangle operators $A_t$ and hexagon operators $B_h$ are defined as:
\begin{equation}
    A_t=\prod_{i\in t}X_i,\ \ \ B_h=\prod_{i\in h}Z_i.
\end{equation}
Namely, triangle operators $A_t$ is the product of $X$ operators on the vertices of each triangle; hexagon operators $B_h$ is the product of $Z$ operators on the vertices of each hexagon. This toric code Hamiltonian for kagome lattice [Eq.~(\ref{Eq:KTC})] can be translated to the usual vertex and plaquette terms by mapping the qubits on vertices of kagome lattice to the links of the root honeycomb lattice.

\subsection{Nearest Neighbor (NN) Ising Coupling}

\begin{figure*}[t]
    \centering
    \includegraphics[width=\textwidth]{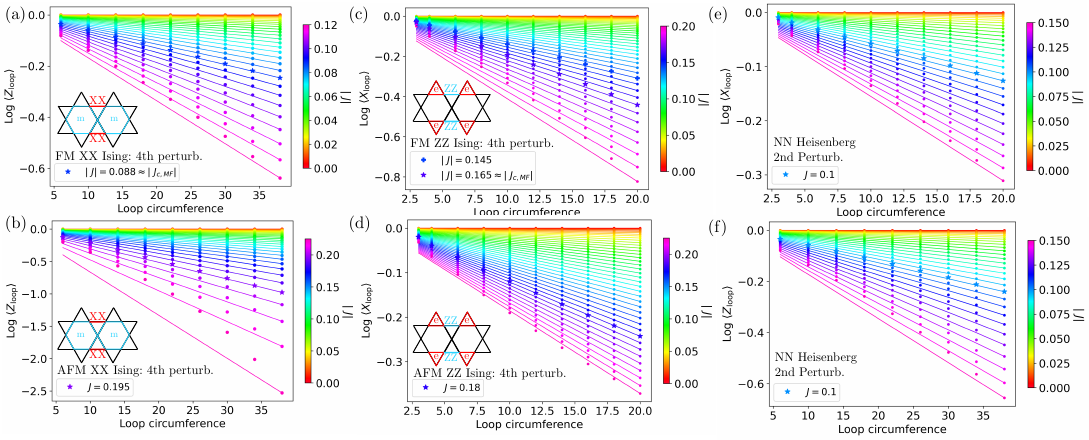}
    \caption{\textbf{Kagome toric code perturbed by NN Ising and Heisenberg coupling} (a-b) Expectation values of $Z_{\text{loop}}$ operators at various NN XX-type Ising coupling strengths. NN XX-type Ising coupling creates a pair of m-anyons residing on the hexagons. (c-d)  Expectation values of $X_{\text{loop}}$ operators at various NN ZZ-type Ising coupling strengths. NN ZZ-type Ising coupling creates a pair of e-anyons residing on the triangles. (e-f) Expectation values of $Z_{\text{loop}}$ and $X_{\text{loop}}$ operators at various NN Heisenberg coupling strengths. }
    \label{fig:KTC_IsingPert_HeisenPert}
\end{figure*}

To begin with, we study the kagome toric code perturbed by nearest neighbor Ising coupling. We considered two types of Ising coupling: NN XX coupling and NN ZZ coupling, see Figure~\ref{fig:KTC_IsingPert_HeisenPert}(a-d).

\textbf{XX-type Ising coupling} Consider NN XX Ising coupling, Figure~\ref{fig:KTC_IsingPert_HeisenPert}(a,b):
\begin{equation}
    H_1=J\sum_{\langle i,j\rangle}X_iX_j
\end{equation}
In this case, all $X$-loop operators remain as good quantum number. We aim to study the properties of local $Z$-loop operators (as products of hexagon operators), Figure~\ref{fig:KagomeTC_AtBh_GLoops_LLoops}(c).

To perform the perturbative calculation, we choose the unperturbed ground state to be stabilized by global $X$-loops $(\Gamma_{X1},\Gamma_{X2})$ as in Figure~\ref{fig:KagomeTC_AtBh_GLoops_LLoops}(b), in addition to the independent triangle and hexagon operators. We should mention that for large system, the choice of the unperturbed ground state should not affect the result of local operators, given the topological nature of model.

We perform fourth order perturbation on a kagome lattice of $10\times 10$ unit cells, namely 300 qubits, with PBC. We compute the expectation value of $Z_{\text{loop}}$ operators of various sizes. The result is summarized in Figure~\ref{fig:KTC_IsingPert_HeisenPert}(a-b).

The NN XX-Ising coupling creates a pair of m-anyons residing on the hexagons, Figure~\ref{fig:KTC_IsingPert_HeisenPert}(a). The hexagons in the kagome lattice forms a triangular lattice. Correspondingly, the perturbed toric code is dual to a transverse field Ising model on triangular lattice:
\begin{equation}
    \tilde{H}_{\triangle-\text{TFIM}}^{\text{XX Ising}}=-\sum_{i}\tau^x_i+2J\sum_{\langle i,j\rangle}\tau^z_{i}\tau^z_{j}
    \label{Eq:DualTFIM_XXIsing}
\end{equation}
where $\tau^{x,z}$ are the Pauli matrices on the sites of the dual triangular lattice and in the equation above, the summation is over the sites on the dual triangular lattice as well. Due to the geometric frustration of the dual triangular lattice, we should differentiate between ferromagnetic ($J<0$) and anti-ferromagnetic ($J>0$) couplings.

On the ferromagnetic side, increasing the coupling strength should drive the system from $\mathbb{Z}_2$ topological order to a state with spontaneous $\mathbb{Z}_2$ symmetry breaking. Indeed, in Figure~\ref{fig:KTC_IsingPert_HeisenPert}(a), we observe a deviation from perimeter when the coupling strength is around $|J|=0.088$, which is quite close to the mean field critical coupling strength $|J_{c,MF}|=1/12$, Figure~\ref{fig:KTC_IsingPert_HeisenPert}(a).

Meanwhile, on the anti-ferromagnetic side, we observe the deviation from perimeter law at a higher coupling strength $J=0.195$, Figure~\ref{fig:KTC_IsingPert_HeisenPert}(b). Qualitatively, this should be due to the geometric frustration of the underlying lattice. Indeed, the AFM coupling drives the system through a transition/crossover from $\mathbb{Z}_2$ topological order with 4-fold ground-state degeneracy with PBC to a state with perhaps macroscopic ground-state degeneracy.

\textbf{ZZ-type Ising coupling} Consider NN ZZ-Ising coupling in Figure~\ref{fig:KTC_IsingPert_HeisenPert}(c-d):
\begin{equation}
    H_1=J\sum_{\langle i,j\rangle}Z_iZ_j
\end{equation}
In this case, all $Z$-loop operators remain as good quantum number. Therefore, we aim to study the properties of local $X$-loop operators (as products of triangle operators), Figure~\ref{fig:KagomeTC_AtBh_GLoops_LLoops}(c).

To perform the perturbative calculation, we choose the unperturbed ground state to be stabilized by global $Z$-loops $(\Gamma_{Z1},\Gamma_{Z2})$ as in Figure~\ref{fig:KagomeTC_AtBh_GLoops_LLoops}(b), in addition to the independent triangle and hexagon operators. We emphasize again that for large system, the choice of the unperturbed ground state should not affect the result of local operators, given the topological nature of model.

We perform fourth order perturbation on a kagome lattice of $10\times 10$ unit cells, namely 300 qubits, with PBC. We compute the expectation value of $X_{\text{loop}}$ operators of various sizes. The result is summarized in Figure~\ref{fig:KTC_IsingPert_HeisenPert}(c-d).

As shown in Figure~\ref{fig:KTC_IsingPert_HeisenPert}(c-d), the NN ZZ-Ising coupling creates a pair of e-anyons residing on the triangles. The triangles in kagome lattice form a honeycomb lattice, with two sublattices. In the current case, two sublattices of the honeycomb lattice decouples. As a result, the kagome toric code perturbed by NN ZZ-Ising is dual to two copies of transverse field Ising model on triangular lattice. The Hamiltonian for each copy is given by:
\begin{equation}
    \tilde{H}_{\triangle-\text{TFIM}}^{\text{ZZ Ising}}=-\sum_{i}\tau^x_i+J\sum_{\langle i,j\rangle}\tau^z_{i}\tau^z_{j}
    \label{Eq:DualTFIM_ZZIsing}
\end{equation}
Again, $\tau^{x,z}$ are the Pauli matrices on the sites of the dual triangular lattice and in the equation above, the summation is over the sites on the dual triangular lattice as well. Also, we should differentiate the ferromagnetic ($J<0$) and anti-ferromagnetic ($J>0$) couplings.

On the side of FM coupling Figure~\ref{fig:KTC_IsingPert_HeisenPert}(c), we observe a deviation from perimeter law at coupling strength $|J|=0.145$, which is quite close the mean field critical point is at $|J_{c,MF}|=1/6$. Notice that in the current case of NN ZZ-Ising coupling, the ``critical'' coupling is about twice of the situation with NN XX-Ising coupling. The difference can be understood from the dual TFIM. Indeed, the dual Ising coupling strength differs by a factor of two, see Eqs.~(\ref{Eq:DualTFIM_XXIsing})(\ref{Eq:DualTFIM_ZZIsing}) and Figure~\ref{fig:KTC_IsingPert_HeisenPert}(a-d).

Meanwhile, on the side of AFM coupling Figure~\ref{fig:KTC_IsingPert_HeisenPert}(d), we observe a deviation from perimeter law at around $J=0.18$. Again, compared to the FM coupling, this is a larger ``critical'' coupling strength.

\subsection{Heisenberg Coupling}

We try to ask the question of how is the kagome toric code state is affected by the NN Heisenberg coupling. Namely, we add NN Heisenberg coupling:
\begin{equation}
    H_1=J\sum_{\langle i,j\rangle}\left(X_iX_j+Y_iY_j+Z_iZ_j\right)
\end{equation}
as a perturbation to the kagome toric code Hamiltonian. We perform \emph{second} order perturbation calculation on kagome lattice of $10\times 10$ unit cells, namely 300 qubits, with PBC. The unperturbed ground state is chosen to be stabilized by $(\Gamma_{X1},\Gamma_{X2})$ in addition to the independent triangle and hexagon operators.


As shown in Figure~\ref{fig:KTC_IsingPert_HeisenPert}(e-f), when the coupling strength is small $J\ll 0.1$, the expectation values of both $X$-loop operators and $Z$-loop operators exhibits perimeter law. Deviation from perimeter law is observed when the coupling strength is large $J\gtrsim0.1$. Notice that both loop operators shows a deviation of perimeter law at roughly the same coupling strength $J\approx 0.1$. A comment is needed: at the second order perturbation, it is not necessary to distinguish between FM ($J<0$) or AFM ($J>0$) couplings. 

This result suggests that both $e$ and $m$ anyons have a tendency toward confinement under the NN Heisenberg perturbation. Hence, the ground state of kagome Heisenberg model is different from the toric code ground state of Eq.~(\ref{Eq:KTC}). While the ground state of strong FM coupling should be a spontaneous symmetry breaking state, the ground state of strong AFM coupling is more nontrivial and interesting.

Indeed, previous studies suggest that the candidate ground states for the AFM kagome Heisenberg model include gapped $\mathbb{Z}_2$ spin liquid \cite{PhysRevB.45.12377,PhysRevB.83.224413,PhysRevB.96.205150,doi:10.1126/science.1201080,PhysRevLett.109.067201,jiang2012TOfromTE} and gapless Dirac $U(1)$ spin liquid \cite{PhysRevB.63.014413,PhysRevLett.98.117205,PhysRevB.87.060405,PhysRevB.89.020407,PhysRevB.91.020402,PhysRevX.7.031020}. While the difference between toric code ground state and the \emph{gapless} Dirac $U(1)$ spin liquid seems obvious (gapped v.s. gapless), the difference from the gapped $\mathbb{Z}_2$ spin liquid is more subtle. It turns out that the crucial difference (for both cases) lies in the realization of symmetry on the anyonic excitations. The anyons in the toric code Hamiltonian Eq.~(\ref{Eq:KTC}) transforms trivially/faithfully under the kagome lattice symmetry. Meanwhile, the anyonic excitations in both candidate states, namely $\mathbb{Z}_2$ and Dirac $U(1)$ spin liquids, transform projectively under the lattice symmetry. In particular, the difference between toric code and the candidate $\mathbb{Z}_2$ spin liquid is mathematically classified by the second group cohomology $\mathcal{H}^2\left(G_s, A_s\right)$ \cite{PhysRevB.96.205150,PhysRevB.97.094422,PhysRevB.97.115138,PhysRevB.100.115147}, where $G_s$ is the global symmetry group and $A_s=\mathbb{Z}_2\otimes\mathbb{Z}_2$ is the Abelian fusion group for the $\mathbb{Z}_2$ topological order. Indeed, the candidate $\mathbb{Z}_2$ spin liquid for AFM kagome Heisenberg model belongs to the symmetry enriched topological order, characterized by nontrivial symmetry fractionalization pattern \cite{PhysRevB.100.115147}. 

To summarize, the perturbative calculation of the $X_{\text{loop}}$ and $Z_{\text{loop}}$ expectation values indicates a tendency toward confinement for both the $e$ and $m$ anyons, which in our toric code starting point are both trivial in terms of symmetry fractionalization patterns. This result indicates the emergence of a qualitatively different ground state at strong NN Heisenberg coupling. However, the detailed nature of the transition is beyond the scope of current study and remains to be explored in the future.

\section{Conclusion and Open Questions}
\label{Sec:Conclusion_discussion}

To conclude, we developed a systematic stabilizer perturbation theory based on the local Schrieffer-Wolff transformation. The stabilizer perturbation theory is constructed to be locality preserving and symmetry preserving. We applied the stabilizer perturbation theory to a few concrete models. Static properties can be computed to high accuracy when the perturbation is weak. In addition, the stabilizer perturbation allows us to estimate the transition point out of the phase represented by the unperturbed stabilizer state. 

We end this paper with a few interesting questions that is worth investigating in the future.

One important question is on performing resummation of the perturbative series. This includes but not limited to the following aspects: (1) Search for alternative (and hopefully better) approximation to the unitary $e^{S}$ other than Taylor expansion \cite{beguvsic2025simulating}; (2) The application of resummation technique, such as Pad\'{e} approximant \cite{PhysRev.124.768}.  A systematic comparison between various resummation technique can be a useful study for further development of stabilizer perturbation theory.

Another natural question to ask is on the generalization to qudit system. Recently, researchers realized that generalized $\mathbb{Z}_N$ toric code in Ref.~\cite{watanabe2023ground} can behave quite differently from the qubit toric code model. It is interesting to generalize the current stabilizer perturbation theory to qudit system and see if the perturbative calculation can provide any insight on the generalized $\mathbb{Z}_N$ toric code model.

Perturbative calculation of entanglement entropy as well as the quantum magic can be an interesting and challenging future direction. Ref.~\cite{PhysRevB.99.205109} constructed perturbative entanglement Hamiltonian based on the first order perturbation. A natural question is whether it is possible to generalize to higher order in perturbation. In addition, it is equally interesting to ask whether certain quantum magic monotone, as a measure of quantum resource \cite{veitch2014resourceStab,bravyi2019simulation}, can be computed perturbatively.

Last but not least, besides static properties, computing dynamical properties, such as dynamical correlation function and time-dependent perturbation theory, is an important future task of the stabilizer perturbation theory. The computation of dynamical properties necessarily involves the description of excited states. Generally, the Schrieffer-Wolff transformation allows a perturbative description of a subspace with some excited state included \cite{BRAVYI20112793}. The technical question is whether Pauli algebra and the stabilizer nature of the unperturbed states can introduce any efficiency in the computation. On the other hand, perturbative simulation of the quantum time evolution (if possible) is also intriguing. The computational resource scales exponentially with the number of non-Clifford gates involved. It would be interesting to explore the limitation of the perturbative simulation and compare with other simulation methods, such as tensor network based method (where the required exponential computational resource comes from the entanglement growth \cite{PAECKEL2019167998}).


\acknowledgements
We are grateful to Shilin Huang for insightful discussion. 
This work is supported by the Hong Kong Research Grants Council (ECS 26308021), the Croucher Foundation (CIA23SC01), and the Fei Chi En Education and Research Fund.
X.Y. acknowledges the support of Hong Kong Research Grants Council through Grant No. PDFS2425-6S02. K.L. is supported by the National University of Singapore start-up grants A-0009991-00-00 and A-0009991-01-00 awarded to Liujun Zou. 

\section*{Data Availability}
The code and data for numerical computation is available upon reasonable request.

\appendix

\section{Numerical Pauli Algebra}
\label{Sec:PauliAlgebra_Implementation}

Based on the binary encoding of an operator $\mathcal{O}$, one can numerically perform the arithmetic of Pauli algebra. Here, we outline our implementation (which works well for the current purpose and is subject to potential further optimization).

\subsubsection{Scalar Multiplication}

Scalar multiplication is given by:
\begin{equation}
    \hat{O}^{\prime}=\lambda\hat{O}=\sum_{i=1}^{N_{\text{op}}}\lambda c_iP_i
\end{equation}
Then, the numerical implementation is straightforward:
\begin{equation}
    \begin{split}
        &\mathsf{CM}_{\hat{O}^{\prime}}=\mathsf{CM}_{\hat{O}}\\
        &\mathsf{Coeff}_{\hat{O}^{\prime}}=\lambda\mathsf{Coeff}_{\hat{O}}=\begin{bmatrix}
            \lambda c_1 & \lambda c_2 & \cdots & \lambda c_{N_{\text{op}}}
        \end{bmatrix}^{\text{T}}
    \end{split}
\end{equation}

\subsubsection{Operator Addition}

Two operators $\hat{O}_{1,2}$ can be added together:
\begin{equation}
\begin{split}
    \hat{O}_3=&\hat{O}_1+\hat{O}_2\\
    =&\sum_{i=1}^{N_{\text{op}1}+N_{\text{op}2}}c_iP_i
\end{split}
\end{equation}
where $\hat{O}_1=\sum_{i=1}^{N_{\text{op}1}}c_iP_i$ and $\hat{O}_2=\sum_{i=N_{\text{op}1}+1}^{N_{\text{op}1}+N_{\text{op}2}}c_iP_i$. In the second line, we did not combine the same terms.

A straightforward numerical implementation of operator summation is as follows:
\begin{equation}
    \begin{split}
        &\mathsf{CM}_{\hat{O}_{3}}=\begin{bmatrix}
            \mathsf{CM}_{\hat{O}_{1}}\\
            \mathsf{CM}_{\hat{O}_{2}}
        \end{bmatrix},\ \ \ 
        \mathsf{Coeff}_{\hat{O}_{3}}=\begin{bmatrix}
            \mathsf{Coeff}_{\hat{O}_{1}} \\ \mathsf{Coeff}_{\hat{O}_{2}}
        \end{bmatrix}
    \end{split}
\end{equation}
The operation above already defines the operator summation. One can perform two more steps of operation - combine the terms with the same Pauli strings and deleting terms with zero coefficients.

\subsubsection{Operator Subtraction}

Operator subtraction can be defined by combining scalar multiplication and operator summation. Indeed,
\begin{equation}
\begin{split}
    \hat{O}_3=&\hat{O}_1-\hat{O}_2
    =\hat{O}_1+ (-1) * \hat{O}_2
\end{split}
\end{equation}
The numerical outcome will be:
\begin{equation}
    \begin{split}
        &\mathsf{CM}_{\hat{O}_{3}}=\begin{bmatrix}
            \mathsf{CM}_{\hat{O}_{1}}\\
            \mathsf{CM}_{\hat{O}_{2}}
        \end{bmatrix},\ \ \ 
        \mathsf{Coeff}_{\hat{O}_{3}}=\begin{bmatrix}
            \mathsf{Coeff}_{\hat{O}_{1}} \\ -\mathsf{Coeff}_{\hat{O}_{2}}
        \end{bmatrix}
    \end{split}
\end{equation}

\subsubsection{Operator Multiplication}

Consider the multiplication of two operators $\hat{O}_{1,2}$:
\begin{equation}
    \hat{O}_l=\sum_{i=1}^{N_{\text{op},l}}c_{l,i}P_{l,i}
\end{equation}
The product of these two operators is given by:
\begin{equation}
    \begin{split}
        \hat{O}_3=&\hat{O}_1\hat{O}_2
        =\sum_{i=1}^{N_{\text{op}1}}\sum_{j=1}^{N_{\text{op}2}}c_{1i}c_{2j}P_{1i}P_{2j}
    \end{split}
\end{equation}

Without combining same terms, the check-matrix $\mathsf{CM}_{\hat{O}_{3}}$ is an $(N_{\text{op}1}*N_{\text{op}2})\times (2N)$ binary matrix and the coefficients $\mathsf{Coeff}_{\hat{O}_{3}}$ is an $(N_{\text{op}1}*N_{\text{op}2})$-component vector.

The rows of the check-matrix $\mathsf{CM}_{\hat{O}_{3}}$ can be labeled by a pair of integers $[ij]$ with $i=1,2,\cdots,N_{\text{op}1}$  and $j=1,2,\cdots,N_{\text{op}2}$. In particular, the $[ij]$-th row is given by the binary summation of $i(j)$-th row of $\mathsf{CM}_{\hat{O}_{1(2)}}^{[i(j)]}$:
\begin{equation}
    \mathsf{CM}_{\hat{O}_{3}}^{[ij]}=\left(\mathsf{CM}_{\hat{O}_{1}}^{[i]}+\mathsf{CM}_{\hat{O}_{2}}^{[j]}\right)\mod2.
\end{equation}
Correspondingly, the components in $\hat{O}_{3,\text{coeff}}$ is also labeled by the same pair of integers $[ij]$, with the $[ij]$-th component given by:
\begin{equation}
    \mathsf{Coeff}_{\hat{O}_{3}}^{[ij]}=\mathsf{Coeff}_{\hat{O}_1}^{[i]}*\mathsf{Coeff}_{\hat{O}_j}^{[j]}*F\left(\mathsf{CM}_{\hat{O}_{1}}^{[i]},\mathsf{CM}_{\hat{O}_{2}}^{[j]}\right)
\end{equation}
where the function $F\left(\mathsf{CM}_{\hat{O}_{1}}^{[i]},\mathsf{CM}_{\hat{O}_{2}}^{[j]}\right)$ is given by Eq.~(\ref{Eq:PauliProdCoeff}).

\subsubsection{Operator Commutator}

Now, we consider the commutator of two operators $\hat{O}_{1,2}$:
\begin{equation}
    \hat{O}_3=[\hat{O}_1,\hat{O}_2]=\sum_{i=1}^{N_{\text{op}1}}\sum_{j=1}^{N_{\text{op}2}}c_{1i}c_{2j}[P_{1i},P_{2j}]
\end{equation}

It is convenient to check commutation relation of Pauli strings $[P_{1i},P_{2j}]$:
\begin{equation}
    M_{[\hat{O}_1,\hat{O}_2]}=\mathsf{CM}_{\hat{O}_{1}}\ \hat{L}\ \mathsf{CM}_{\hat{O}_{2}}^{\text{T}}\mod2
\end{equation}
The matrix elements of $M_{[\hat{O}_1,\hat{O}_2]}$ dictates the commutation relation:
\begin{equation}
    M_{[\hat{O}_1,\hat{O}_2]}^{(i,j)}=\left\{\begin{split}
        &0\ \ \ \ \text{if}\ [P_{1i},P_{2j}]=0\\
        &1\ \ \ \ \text{if}\ [P_{1i},P_{2j}]\neq 0
    \end{split}\right.
\end{equation}
Suppose the total number of nonzero elements in $M_{[\hat{O}_1,\hat{O}_2]}$ is $N_{[\hat{O}_1,\hat{O}_2]}$. The check matrix $\mathsf{CM}_{\hat{O}_{3}}$ is an $N_{[\hat{O}_1,\hat{O}_2]}\times (2N)$ matrix and the coefficient vector $\mathsf{Coeff}_{\hat{O}_{3}}$ contains $N_{[\hat{O}_1,\hat{O}_2]}$ components.

The rows of the check matrix $\mathsf{CM}_{\hat{O}_{3}}$ is labeled by a pair of integers $[ij]$ where $M_{[\hat{O}_1,\hat{O}_2]}^{(i,j)}=1$. The $[ij]$-th row is given by:
\begin{equation}
    \mathsf{CM}_{\hat{O}_{3}}^{[ij]}=\left(\mathsf{CM}_{\hat{O}_{1}}^{[i]}+\mathsf{CM}_{\hat{O}_{2}}^{[j]}\right)\mod2
\end{equation}
Correspondingly, the components in $\mathsf{Coeff}_{\hat{O}_{3}}$ is also labeled by the same pair of integers $[ij]$, with the $[ij]$-th component given by:
\begin{equation}
    \mathsf{Coeff}_{\hat{O}_{3}}^{[ij]}=2*\mathsf{Coeff}_{\hat{O}_1}^{[i]}*\mathsf{Coeff}_{\hat{O}_j}^{[j]}*F\left(\mathsf{CM}_{\hat{O}_{1}}^{[i]},\mathsf{CM}_{\hat{O}_{2}}^{[j]}\right)
\end{equation}
where the function $F\left(\mathsf{CM}_{\hat{O}_{1}}^{[i]},\mathsf{CM}_{\hat{O}_{2}}^{[j]}\right)$ is given by Eq.~(\ref{Eq:PauliProdCoeff}).

\subsubsection{Employing Translation Invariance}

In many cases, we need to deal with the Pauli algebra for operators with translation invariance. Consider the following translation symmetric operator:
\begin{equation}
    \hat{O}_{\text{TSym}}=\sum_n\left(\hat{T}^{\dagger}\right)^n\hat{O}\hat{T}^n
\end{equation}
Numerically, it is enough to store the check matrix and the coefficients of $\hat{O}$ as well as how translation operator acts on $\hat{O}$:
\begin{center}
    \begin{tabular}{c c c c}
    \hline
        Trans. Inv. Op. & Check Matrix & \ \ Coeff. & \ \ Translation \\
        $\hat{O}_{\text{TSym}}=\sum_n\left(\hat{T}^{\dagger}\right)^n\hat{O}\hat{T}^n$ & $\mathsf{CM}_{\hat{O}}$ & $\mathsf{Coeff}_{\hat{O}}$ & $\hat{T}$\\
         & & &\\
        \hline
    \end{tabular}
\end{center}
where the information of translation $\hat{T}$ dictates how the elements in the check matrix $\mathsf{CM}_{\hat{O}}$ shift under translation operation.

The addition of two translation invariant operator is straightforward to perform:
\begin{equation}
    \hat{O}_{\text{TSym},1}+\hat{O}_{\text{TSym},2}=\sum_n\left(\hat{T}^{\dagger}\right)^n\left(\hat{O}_1+\hat{O}_2\right)\hat{T}^n
\end{equation}
Namely, it is enough to compute $\hat{O}_1+\hat{O}_2$. Subtraction can be similarly defined
\begin{equation}
    \hat{O}_{\text{TSym},1}-\hat{O}_{\text{TSym},2}=\sum_n\left(\hat{T}^{\dagger}\right)^n\left(\hat{O}_1-\hat{O}_2\right)\hat{T}^n.
\end{equation}

The multiplication is slightly more complicated:
\begin{equation}
    \begin{split}
        \hat{O}_{\text{TSym},1}\hat{O}_{\text{TSym},2}=&\sum_{n_1n_2}\left(\hat{T}^{\dagger}\right)^{n_1}\hat{O}_1\hat{T}^{n_1}\left(\hat{T}^{\dagger}\right)^{n_2}\hat{O}_2\hat{T}^{n_2}\\
        =&\sum_{n_1}\left(\hat{T}^{\dagger}\right)^{n_1}\hat{O}_1\sum_{n_2}\left(\hat{T}^{\dagger}\right)^{n_2-n_1}\hat{O}_2\hat{T}^{n_2-n_1}\hat{T}^{n_1}\\
        =&\sum_{n_1}\left(\hat{T}^{\dagger}\right)^{n_1}\hat{O}_1\hat{O}_{\text{TSym,2}}\hat{T}^{n_1}
    \end{split}
\end{equation}
Instead of computing the full multiplication, one can just compute $\hat{O}_1\hat{O}_{\text{TSym,2}}$.

Similarly, the commutator of two translation invariant operators can be computed as:
\begin{equation}
    \left[\hat{O}_{\text{TSym},1},\hat{O}_{\text{TSym},2}\right]=\sum_{n_1}\left(\hat{T}^{\dagger}\right)^{n_1}\left[\hat{O}_1,\hat{O}_{\text{TSym,2}}\right]\hat{T}^{n_1}.
\end{equation}
It is enough to compute $\left[\hat{O}_1,\hat{O}_{\text{TSym,2}}\right]$ to deduce the full commutator.

\section{Verify the check matrix for de-stabilizer operators}
\label{Sec:Verify_DeStab_CM}

Now, we demonstrate the construction of the check matrix for the de-stabilizer operators  Eq.~(\ref{Eq:DeStab_CM_Result}). Start from the Smith normal form in Eq.~(\ref{Eq:SmithNF_CM_gs}). Since $P$ and $Q$ are invertible, we can perform a few lines of mathematical manipulations:
\begin{equation}
    \begin{split}
        \mathsf{CM}_{\text{gs}}=&P^{-1}\begin{bmatrix}
        \mathbb{I}_{N\times N} & 0_{N\times N}
    \end{bmatrix}Q^{-1}\\
    =&\begin{bmatrix}
        \mathbb{I}_{N\times N} & 0_{N\times N}
    \end{bmatrix}\begin{bmatrix}
        P^{-1} & 0\\
        0 & \tilde{P}^{-1}
    \end{bmatrix}Q^{-1}
    \end{split}
\end{equation}
for some invertible binary matrix $\tilde{P}$. Then, the whole equation above can be rewritten as:
\begin{equation}
   \mathsf{CM}_{\text{gs}}\ Q \begin{bmatrix}
        P & 0\\
        0 & \tilde{P}
    \end{bmatrix}=\begin{bmatrix}
        \mathbb{I}_{N\times N} & 0_{N\times N}
    \end{bmatrix}
\end{equation}
Substituting in the block form of $Q$, we obtain:
\begin{equation}
   \mathsf{CM}_{\text{gs}}\  \begin{bmatrix}
        Q_{11} P & Q_{12} \tilde{P}\\
        Q_{21} P & Q_{22} \tilde{P}
    \end{bmatrix}=\begin{bmatrix}
        \mathbb{I}_{N\times N} & 0_{N\times N}
    \end{bmatrix}
\end{equation}
Hence,
\begin{equation}
    \mathsf{CM}_{\text{gs}}\begin{bmatrix}
        Q_{11} P \\
        Q_{21} P 
    \end{bmatrix}=
        \mathbb{I}_{N\times N} 
\end{equation}
This equation is equivalent to:
\begin{equation}
    \begin{bmatrix}
        \left(Q_{21}P\right)^{\text{T}} & \left(Q_{11}P\right)^{\text{T}}
    \end{bmatrix}\ \hat{L}\ \mathsf{CM}_{\text{gs}}^{\text{T}}=\mathbb{I}_{N\times N}.
\end{equation}
And the check matrix for de-stabilizer operators $\mathsf{CM}_{\text{dgs}}$ [Eq.~(\ref{Eq:DeStab_CM_Result})] follows. Lastly, the choice of $\tilde{P}$ is not important. This is because of the following relation:
\begin{equation}
    \mathsf{CM}_{\text{gs}}\begin{bmatrix}
        Q_{12} \\
        Q_{22} 
    \end{bmatrix}=
        0_{N\times N} 
\end{equation}
This relation follows from the construction of the Smith normal form.




\bibliography{StabilizerPerturbation}

\end{document}